\documentclass[10pt,preprint]{aastex}

\shorttitle{ROSAT/SDSS AGN: DR5 Sample}
\shortauthors{Anderson, Scott F. et al.}

\begin{document}

\title{A Large, Uniform Sample of X-ray Emitting AGN from the ROSAT 
All-Sky and Sloan Digital Sky Surveys: the Data Release~5 Sample}

\author{Scott F. Anderson\altaffilmark{1},
Bruce Margon\altaffilmark{2},
Wolfgang Voges\altaffilmark{3},
Richard M. Plotkin\altaffilmark{1},
David Syphers\altaffilmark{1},
Daryl Haggard\altaffilmark{1},
Matthew J. Collinge\altaffilmark{4},
Jillian Meyer\altaffilmark{1},
Michael~A.~Strauss\altaffilmark{4}, 
Marcel A. Ag\"ueros\altaffilmark{1},
Patrick B. Hall\altaffilmark{5},
L.~Homer\altaffilmark{1},
\v Zeljko~Ivezi\'c\altaffilmark{1},
Gordon~T.~Richards\altaffilmark{6},
Michael W. Richmond\altaffilmark{7},
Donald~P.~Schneider\altaffilmark{8},
Gregory Stinson\altaffilmark{1},
Daniel~E.~Vanden~Berk\altaffilmark{8},
Donald~G.~York\altaffilmark{9}
}

\email{anderson@astro.washington.edu, margon@stsci.edu, 
wvoges@mpe.mpg.de, plotkin@astro.washington.edu}

\altaffiltext{1}{University of Washington, Department of
   Astronomy, Box 351580, Seattle, WA 98195}

\altaffiltext{2}{Space Science Telescope Institute, 3700 San
Martin Drive, Baltimore, MD, 21218}

\altaffiltext{3}{Max Planck-Institute f\"ur extraterrestrische
Physik, Geissenbachstr. 1, D-85741 Garching, Germany}

\altaffiltext{4}{Princeton University Observatory, Princeton,
NJ 08544}

\altaffiltext{5}{Department of Physics \& Astronomy, York
        University, 4700 Keele St., Toronto, ON, M3J 1P3, Canada}

\altaffiltext{6}{Department of Physics and Astronomy,
The Johns Hopkins University, 3400 North Charles Street,
Baltimore, MD 21218-2686}

\altaffiltext{7}{Department of Physics,
Rochester Institute of Technology, Rochester, NY 14623-5603}

\altaffiltext{8}{Department of Astronomy and 
Astrophysics, The
   Pennsylvania State University, University Park, PA 16802}

\altaffiltext{9}{Astronomy and Astrophysics Center, University 
of Chicago, 5640 South Ellis Avenue, Chicago, IL 60637}

\begin{abstract}

We describe further results of a program aimed to yield 
$\sim$10$^4$ fully characterized optical identifications of ROSAT X-ray 
sources. Our program employs X-ray data from the ROSAT All-Sky 
Survey (RASS), and both optical imaging and spectroscopic data 
from the Sloan Digital Sky Survey (SDSS). RASS/SDSS data from 5740~deg$^2$ 
of sky spectroscopically covered in SDSS Data Release 5 (DR5) provide an 
expanded catalog of 7000 confirmed quasars and other AGN that 
are probable RASS identifications. Again in our expanded catalog, the 
identifications as X-ray sources are statistically secure, with only a few 
percent of the SDSS AGN likely to be randomly superposed on unrelated RASS 
X-ray sources. Most identifications continue to be quasars and Seyfert 1s 
with $15<m<21$ and $0.01<z<4$; but the total sample size has grown to 
include very substantial numbers of even quite rare AGN, e.g., 
now including several hundreds of candidate X-ray emitting BL Lacs and 
narrow-line Seyfert 1 galaxies.
In addition to exploring rare subpopulations, such a large total sample 
may be useful when considering
correlations between the X-ray and the optical, and may also serve as a 
resource list from which to select the ``best" object (e.g., X-ray 
brightest AGN of a certain subclass, at a preferred redshift or 
luminosity) for follow-on X-ray spectral or alternate detailed studies.

\end{abstract}

\newpage
\section{Introduction}

The premier X-ray imaging survey of the astronomical sky is the ROSAT 
All-Sky Survey (hereafter RASS; Voges et al. 1999, 2000). In the 1990s, 
RASS surveyed most of the celestial sphere in the 
$0.1-2.4$~keV 
range with the Position Sensitive Proportional Counter (PSPC; Pfeffermann 
et al. 1988) to a typical limiting sensitivity of $\sim 
10^{-13}$~erg~cm$^{-2}$~s$^{-1}$, cataloging more than $10^5$ X-ray 
sources in the RASS Bright and Faint Source Catalogs.

With RASS's typical X-ray positional uncertainty of order $\sim10-30''$, 
the effort involved in optically identifying a large fraction of the 
cataloged abundant RASS sources may appear daunting; and yet, 
several large-scale identification efforts are currently making 
substantial progress. Thousands of optical identifications for RASS 
sources are now suggested, albeit with various levels of identification 
confidence on the nature of the counterparts, ranging from: statistical 
cross-correlations with Schmidt 
photographic plate images (e.g., see radio/optical/X-ray matches cataloged 
in Flesch and Hardcastle 2004); to optical spectral identification 
estimates from low spectral-resolution photographic Schmidt objective 
prism plates (e.g., see 
Zickgraf et al. 2003; Mickaelian et al. 2006); to digital optical 
photometry and
multi-object spectrophotometry (e.g., see RASS Bright and Faint Source 
Catalog identifications in Anderson et al. 2003).
 
For comparison to such recent RASS efforts involving thousands of 
suggested optical identifications each, it may be recalled that the most 
complete optical identification effort at similar depth from {\it 
Einstein} X-ray 
data was the Extended Medium Sensitivity Survey (EMSS; e.g., Gioia 
et al. 1984, Stocke et al. 1991). Quasars/AGN were the predominant class 
in the EMSS, accounting for about half of the $\sim800$ {\it Einstein} 
identifications. Although the EMSS total of $\sim400$ AGN identifications 
is quite large in 
aggregate,  certain subsets are critically small when subdivided; for 
example, although the EMSS provided fundamental insights into BL Lac
evolution (Morris et 
al. 1991), the entire EMSS sample included only $\sim$40 BL~Lacs.

In order to substantively expand on such earlier {\it Einstein} 
programs, current efforts should provide samples  
significantly exceeding $\sim10^3$ spectroscopically secure optical 
identifications. Our approach, detailed previously in Anderson et al. 2003 
(hereafter, Paper 1), relies on the very good sensitivity match between 
AGN detectable in X-rays with RASS and AGN accessible to good quality 
optical spectra with the Sloan Digital Sky Survey (SDSS; York et al. 
2000). Our ultimate 
aim has been the cataloging and characterization of $\sim10^4$ X-ray 
identifications; the AGN in this sample are not only
statistically secure in 
their identifications as X-ray sources, but are also accompanied by 
high quality and uniform data in the X-ray from RASS and in the optical 
(both photometry and spectroscopy) from SDSS.

In section 2, we briefly 
review relevant aspects of the SDSS data, and
the selection and confirmation of candidate AGN 
optical counterparts from SDSS; detailed descriptions of these 
aspects, and a much more complete set of references, may be found 
in Paper 1. In section 3, we present results from our 
RASS/SDSS program for 5740~deg$^2$ of sky (about $4\times$ the area
coverage considered in Paper 1), providing an updated
and expanded catalog of X-ray and optical
properties of 7000 likely SDSS AGN
identifications for RASS X-ray sources.
In sections 3-4, we discuss some AGN
subclasses of special interest, including more than 260 X-ray emitting BL 
Lac 
candidates,
and 770 candidate X-ray emitting narrow-line Seyfert 1s and related 
quasars.
Section 5  discusses the reliability of the 
identifications
and some ensemble properties of the expanded sample. Section~6 provides
an updated example optical/X-ray correlation study.
A short summary is provided in Section~7. 

\section{Selected Aspects of SDSS Relevant to RASS Identifications}

The SDSS is 
creating an optical digital imaging and spectroscopic database of
a large portion of the celestial sphere, primarily in a region approaching
$\sim10^4$~deg$^2$  centered on the north Galactic polar cap.
The optical data are obtained by a dedicated 2.5m telescope,
located at Apache Point Observatory, New Mexico,
equipped with a large-format mosaic camera that can image
$\sim10^2$~deg$^2$ in 5 colors ($u,g,r,i,z$) in a single night, as well
as a multifiber spectrograph which obtains the spectra of 640 objects
within a 7~deg$^2$ field simultaneously.  The imaging database
is used to select objects for the SDSS spectroscopic survey, which
includes ($\lambda/\Delta\lambda\sim1800$)
spectrophotometry covering a broad (3800-9200\AA) wavelength range
for $10^6$ galaxies, $10^5$ quasars, and $10^5$ stars.
Details on SDSS hardware, software, astrometric, photometric,
and spectral data may be found in a variety of papers, including
Fukugita et al. (1996), Gunn et al. (1998), 
Lupton et al. (1999), 
York et al. (2000), Hogg et al. (2001), Stoughton et al.
(2002), Smith et al. (2002), Pier et al. (2003), Ivezi\'c et al.
(2004), Gunn et al. (2006), 
and Tucker et al. (2006).
A description of the previous SDSS Public Data Release (DR4) is given
by Adelman-McCarthy et al. (2006).

The depths of RASS and SDSS are well-matched.
For example, quasars and BL~Lacs are known to 
have extreme X-ray to optical flux ratios 
of order $log~(f_x/f_{opt})\sim 1.0-1.5$ 
(Stocke et al. 1991); so even unusually 
faint RASS optical counterparts will have magnitudes 
brighter than $m<20-21$. The SDSS imaging survey at this depth provides 
accurate colors and magnitudes 
for nearly all RASS counterparts, and SDSS multi-object
followup spectroscopy also typically
yields excellent quality data.
The RASS/SDSS area is also covered by the 
NVSS and/or
FIRST 20~cm radio surveys (Condon et al. 1998; Becker, White \& Helfand 
1995), providing radio information for nearly all
RASS/SDSS objects. 

A detailed description of the SDSS ``target selection pipeline" 
(Stoughton et al. 2002) algorithms applicable to RASS sources is provided
in our Paper 1, so here we provide an abbreviated summary.
Optical objects in the SDSS photometric catalogs are automatically 
cross-correlated with X-ray sources in the RASS Bright and 
Faint Source catalogs, and those
SDSS optical objects within $1'$ of the X-ray source positions are 
initially flagged 
as  potential  positional matches to be considered further. Due to 
limitations 
on the number of spare SDSS fibers available for ROSAT identifications, 
and a
$55''$ restriction on the minimum  separation of adjacent fibers, we 
especially favor confirming optical spectroscopy of a single, 
high-priority SDSS candidate counterpart within $27.5''$ 
of the RASS X-ray position for each of the 
$\sim10^4$ highest-significance RASS sources (X-ray detection 
likelihood $\ge10$) that lie within the joint RASS/SDSS sky coverage.
SDSS spectroscopic limits for RASS candidate counterparts
are approximately in the range 
$15<m<20.5$, where $m$ refers here to $g$, or $r$, or $i$ passbands.
 
Each SDSS optical object imaged in a
RASS error circle is assigned to a graded (A,B,C,D,E) priority bin
for follow-on SDSS optical spectroscopy. The priority bins are based on 
typical ratios of X-ray to  optical flux for known subclasses 
of X-ray 
emitters, 
with SDSS optical magnitudes, colors, and image 
morphology, as well as FIRST radio data, serving as indicators
of the likely object class 
during spectroscopic target-selection.
The `ROSAT\_A' and `ROSAT\_B' (two highest priority) categories
are specifically aimed at AGN, and encompass about 
92\% of the identifications presented here. These two categories are also 
highly efficient, with 85\% of all targeted `ROSAT\_A' and `ROSAT\_B' 
objects returning spectra of AGN. `ROSAT\_A' objects are triple 
positional coincidences of a RASS X-ray source, an SDSS optical object, 
and a FIRST radio source, especially aimed at BL~Lacs (and other 
X-ray/radio-emitting quasars). 
`ROSAT\_B' is assigned to SDSS objects
having unusual optical colors indicative of AGN/quasars (Richards et al.
2002);
UV excess and more sophisticated multicolor approaches are consulted,
e.g., see Figure 1. The reader is referred to Paper 1 for 
details on the other ROSAT
target selection categories.

\begin{figure}
\plotone{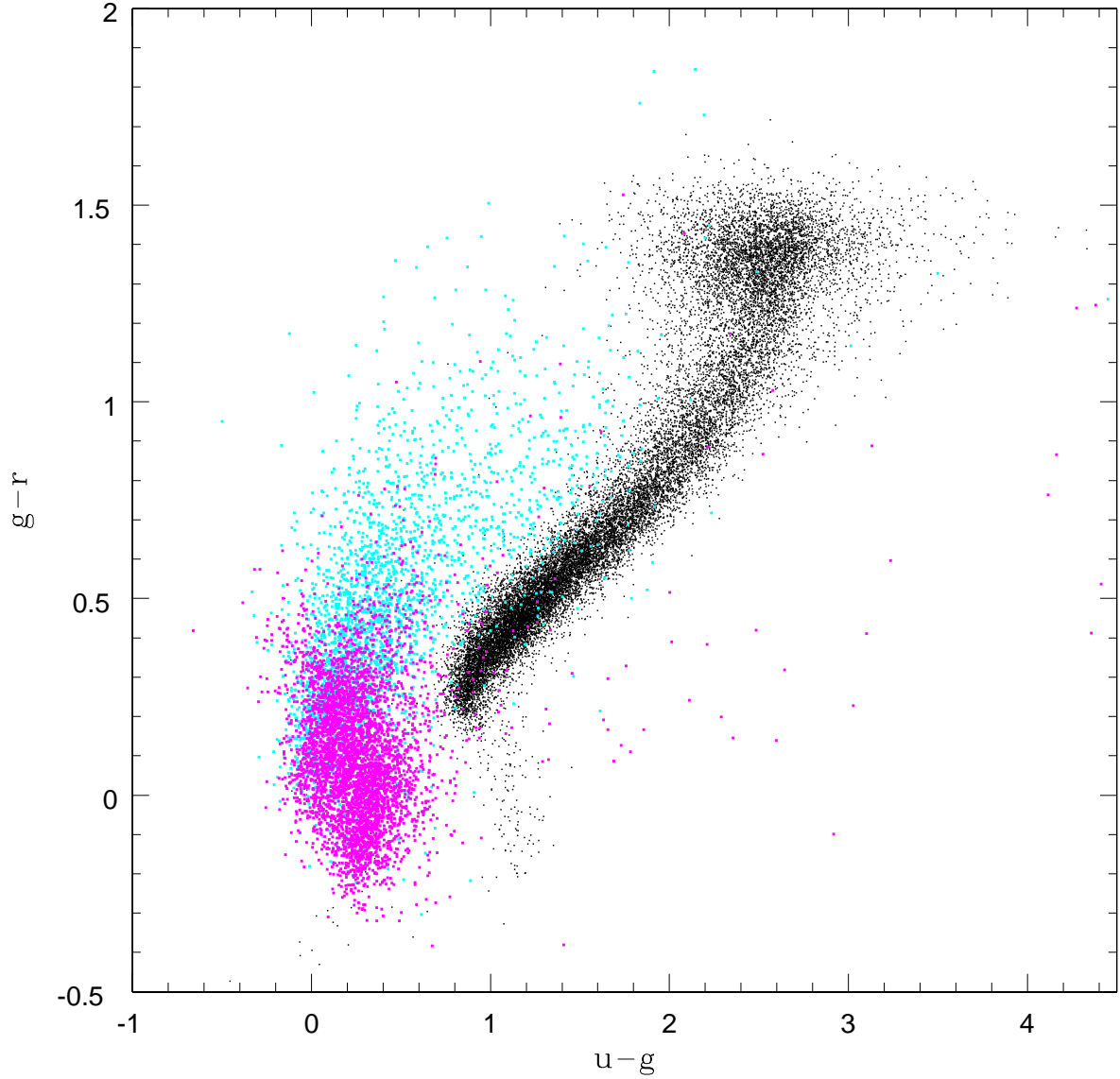}
\figcaption{
Optical SDSS colors of 7000 spectroscopically-confirmed quasars and 
other X-ray emitting AGN in our updated sample are overplotted for 
comparison on the locus of 10,000 anonymous SDSS stellar objects 
(black points). The {\it magenta} points show SDSS colors of 
quasi-stellar X-ray identifications that are unresolved in SDSS 
images, while {\it cyan} points show colors of AGN morphologically 
resolved in SDSS. Note the good color separation of the large 
majority of the confirmed quasar/AGN identifications.
\label{fig1}}
\end{figure}

Additionally, SDSS
objects that are part of the main optical spectroscopic surveys of
$10^{6}$ galaxies 
and $10^{5}$ quasars are assigned their
spectroscopic fibers first, even before the SDSS/RASS target
selection algorithms (making use of spare fibers) come into play. As a 
bonus to this identification program, SDSS spectra are thus 
available for the 
vast majority of such ``tiled targets" (Blanton et al. 2003)
falling fortuitously in RASS error circles; these 
include most
galaxies to Petrosian $r<17.77$, 
SDSS objects to $i<19.1$ with quasar colors,
and FIRST radio sources with 
optical PSF-morphology to $i<19.1$.
On the other hand,
a galaxy targeted as 
part of the SDSS galaxy redshift-survey sample, and merely falling by 
chance  near a RASS error circle, may sometimes eliminate a
desired ROSAT target from any chance of receiving a 
spectroscopic fiber.

Thus, our expanded RASS/SDSS catalog
is not a ``complete" sample, though
in practice we  expect
the catalog may eventually
prove to be reasonably complete 
for most classes of
X-ray emitting quasars and AGN with $15<m<19$.
For these objects, quasar target selection completeness is confirmed to be 
high
(Vanden Berk et al. 2005), and
these may be selected for SDSS spectra by either spare-fiber ROSAT 
or tiled-fiber quasar target selection categories.

\section{Expanded Catalog of 7000 X-ray Emitting AGN From RASS/SDSS}

The expanded catalog of RASS/SDSS
identifications is presented in this section.
The emphasis in this installment of the
RASS/SDSS catalog is on a set of
7000 X-ray emitting quasars/AGN. 
About 1200 of these constituted our
original installment of the RASS/SDSS AGN catalog presented
in Paper 1. For the full sample of 7000
AGN, the identifications are statistically secure, as discussed
in \S 5.

We have examined an early version of the Data Release 5 (DR5) database and 
identified
15,129 SDSS spectra that fall within 1$'$ of a RASS source.  All of these
spectra have been examined by a variety of methods, both algorithmic
and visual, to search 
for plausible AGN identifications for the corresponding RASS X-ray 
sources. Our examination of the SDSS spectra taken within 
RASS error circles emphasizes 
searching especially for the following sets of AGN: (i) 
classic AGN/quasars with strong, broad emission lines; (ii) AGN with 
narrower line 
components; and, (iii) BL Lacs with weak or absent spectral features.

\subsection{Quasars and Other AGN with Strong, Broad Permitted Lines}

In Tables 1 and 2, we present an updated
catalog of spectroscopically confirmed RASS/SDSS quasars and 
closely-related AGN having strong, broad permitted emission lines. All are 
in SDSS DR5 and have optical positions within $1'$ of RASS X-ray sources. 
Sample versions of Tables 1 and 2, listing only the first five entries, 
are included within this paper; the full tables are available 
electronically from the journal. We present the entire updated RASS/SDSS 
AGN catalog, including both new objects and those presented 
previously in Paper 1 (or elsewhere in the literature). Paper 1 included
optical data from very early-on in the SDSS program (e.g., extending
back to the ``Early Data Release" before SDSS photometric calibrations 
were
complete; Stoughton et al. 2002), so here we catalog all 
counterparts with their uniform SDSS DR5 photometric and 
spectroscopic 
parameters.

Included under the category of ``broad-line" 
quasars, are RASS/SDSS objects whose spectra show 
characteristic 
strong 
optical emission lines of AGN, with broad permitted emission having 
velocity width in excess of 1000~km~s$^{-1}$ FWHM; the latter value is 
also 
that used in the DR3 quasar catalog of Schneider et al. (2005), and is close 
to the value of 1200~km~s$^{-1}$ that separates a
bimodal line-width distribution seen in SDSS Seyfert galaxies
(Hao et al. 2005). 
In order to identify such broad-lined quasars, we first considered 
those spectra in which the SDSS
spectroscopic 1D pipeline software (Stoughton et al. 2002) returned a 
best-fit Gaussian FWHM exceeding 1000~km~s$^{-1}$ for any of the following 
emission lines: Ly$\alpha$, NV~1240, SiIV~1400 (blend), CIV~1549, 
CIII]~1909, 
MgII~2800, H$\gamma$, H$\beta$, or H$\alpha$. Of the 15,129 SDSS DR5 
spectra taken in RASS error circles, 7220 spectra appeared to satisfy this 
initial pipeline-based emission line-width criteria within the DR5 
database. An 
additional 17 spectra were found in a subsequent by-eye examination 
(discussed in section 3.2) that are also clearly predominantly 
broad-lined  AGN, but which failed the algorithmic cut discussed above 
based on spectroscopic 
pipeline measures; for example,
this group includes some broad absorption line QSOs (BALQSOs), some cases 
with spectral reduction
problems, etc.

We visually re-examined all 7237 spectra to confirm their nature,
as well as to verify the
pipeline redshifts. Among these, 6582 were
visually confirmed to indeed be spectra of broad-line 
quasars, with only 9 requiring significant redshift corrections to 
the pipeline estimates.
From the surviving list of 6582 confirmed broad-line AGN spectra,
there are 6224
distinct objects; that is, about 5\% of these AGN have more 
than one SDSS spectrum available in DR5.
Although the line-width selection in this paper is based 
almost entirely on SDSS pipeline
Gaussian-fit measures, 99\% of the AGNs cataloged in Paper 1
(where classifications were based on manual line measures) are also 
recovered in this paper; the substantial recovery fraction is
not a surprise of course, but worthy of verification given our far heavier 
reliance in the current paper on SDSS pipeline spectroscopic measures.
Note also that we avoid any
luminosity cuts, and so Tables 1 and 2 include not just most 
classic quasars and Sy~1s,
but also many Sy~1.5s to Sy~1.9s, and even many rare objects 
such as narrow lined Seyfert 1s (NLS1s). 

All AGN have confirming, high-quality SDSS 
optical spectra and imaging, uniformly reduced as part of
DR5. See Figure~2 for representative 
examples of their SDSS spectra. 

\begin{figure}
\plotone{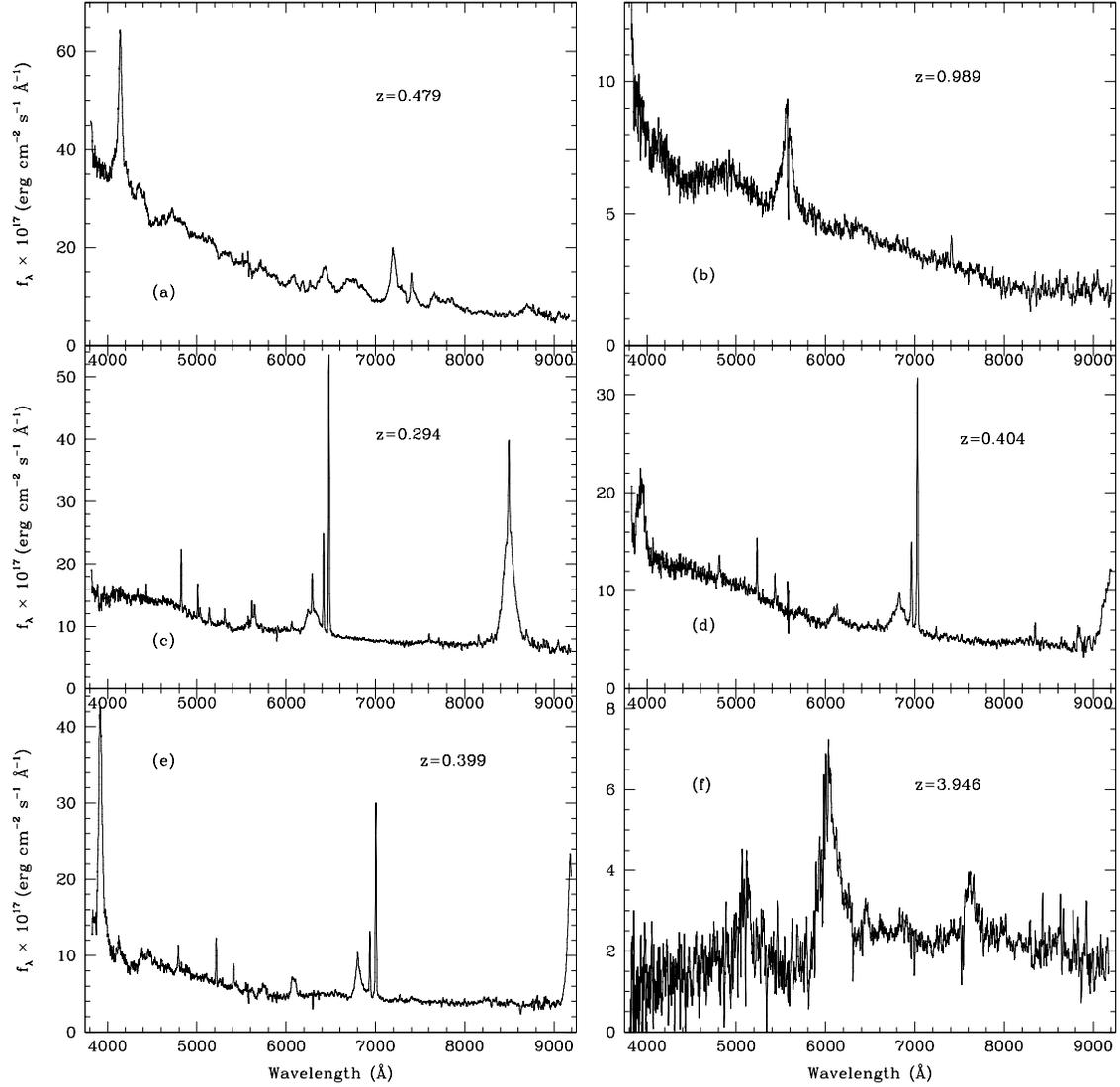}
\caption{
Representative SDSS optical spectra for
RASS/SDSS X-ray emitting quasars/AGN with
broad permitted emission lines, as discussed in section 3.1.
Shown (a-e) are spectra smoothed with a 7 point boxcar for the first 5
objects
listed in Tables 1-2,
along with a high-redshift case (f) reflecting the very broad
range
of redshifts sampled.
All 6224 broad line AGN cataloged
have similar high-quality SDSS spectroscopy.
(a) SDSS J000011.96+000225.3; (b) SDSS J000024.02+152005.4;
(c) SDSS J000102.18$-$102326.9; (d) SDSS J000116.00+141123.0;
(e) SDSS J000132.83+145608.0; (f) SDSS J081009.95+384757.0.
\label{fig2}}
\end{figure}

Following our format in Paper 1, Table~1 includes mainly empirical
characteristics of the 6224 X-ray-emitting RASS/SDSS broad-line 
quasar/AGN 
counterparts, while Table~2 provides mainly derived information. 
Table~1 is ordered according to the RA (J2000) of the RASS X-ray source;
the {\it 1st column} lists the X-ray position,
using RA/Dec nomenclature. The {\it 2nd column} provides the 
optical position/nomenclature (J2000) of the suggested SDSS quasar/AGN
counterpart.
The {\it 3rd through 7th columns} provide uniform DR5 optical PSF
photometry in the 5 SDSS passbands (e.g., Fukugita et al. 1996) in the
{\it asinh} AB system (Lupton, Gunn, \& Szalay 1999).
The {\it 8th column} provides the value of the SDSS imaging 
morphology parameter:
type=6 indicates stellar/unresolved optical morphology, while
type=3 indicates an extended/resolved (i.e., galaxy) morphology.
The {\it 9th column} provides the redshift
measured from SDSS spectra.
The remaining columns of Table~1 emphasize empirical X-ray data 
from the RASS catalogs (e.g., see Voges et al. 1999, 2000). 
The {\it 10th column} provides the RASS X-ray source
count rate (counts~s$^{-1}$) in the 0.1-2.4~keV broadband, 
corrected for 
vignetting. The {\it 11th column} gives the RASS exposure time in seconds.
The {\it 12th and 13th columns} provide
X-ray hardness ratios.
The {\it 14th column} is the X-ray source detection likelihood.
The {\it 15th column}
gives the observed X-ray flux in the 0.1-2.4~keV band; we used
the PIMMS (Portable, Interactive, Multi-Mission Simulator) 
software to convert RASS count rates 
into X-ray fluxes, assuming
a power-law X-ray spectrum with energy index $\alpha_x=1.5$, typical of
low redshift quasars in the RASS PSPC bandpass (e.g., see Schartel 
et al. 1996).

In Table 2, we present further catalog information on the 6224
broad line RASS/SDSS AGN,
emphasizing derived quantities.
For cross reference to Table 1, we repeat
in the
{\it 1st column} and {\it 2nd column},
respectively, the RASS X-ray source and optical counterpart
name/position.
The {\it 3rd column} provides the 
$g$-band PSF magnitude but here 
extinction corrected according to the reddening
maps of Schlegel, Finkbeiner, \& Davis~(1998). 
The {\it 4th column}
repeats the SDSS spectroscopic redshift.
The {\it 5th column} is the X-ray flux (units of
$10^{-13}$~erg~s$^{-1}$~cm$^{-2}$) in the 0.1-2.4~keV band, now
corrected for absorption within the Galaxy,
with X-ray absorbing columns estimated using the $N_H$ column 
density
measures of the Stark et al. (1992) 21cm maps.
The {\it 6th, 7th, and 8th 
columns} give the logarithms of, respectively, inferred broadband
(0.1-2.4~kev) X-ray luminosity (units of erg~s$^{-1}$), 
UV/optical 
monochromatic
luminosity (cgs units of erg~s$^{-1}$~Hz$^{-1}$) at a frequency 
corresponding to rest frame 2500~\AA\ , and  monochromatic
X-ray luminosity (cgs) at 2~kev; we adopt 
values of $H_0$=70~km~s$^{-1}$~Mpc$^{-1}$, $\Omega_M=0.3$, and
$\Omega_{\Lambda}=0.7$ for deriving the luminosities in Table~2.
In converting from corrected broadband (X-ray 0.1--2.4~kev, and optical
$g$-band) fluxes to luminosities, we again
assume an X-ray power-law spectrum with energy index $\alpha_x=1.5$,
and  an optical power-law with energy index $\alpha_o=0.5$.
The {\it 9th column} lists $\alpha_{ox}$, the 
slope of a hypothetical power law in energy from the UV/optical to X-ray 
(i.e., connecting 2500~\AA\ and 2~kev). 
The {\it 10th column} provides brief
comments, for example noting selected objects that are radio sources,
listing an alternate name (shortened, if needed, to the first 20 
characters)
taken from NED for about 10\% of the quasars,
and noting the $\sim$1.5\% of the cases where two of our cataloged AGN 
fall within the same RASS error circle.
(The latter are denoted by `ambigID' in the comments, and eight such 
cases
involve AGN pairs at similar redshifts).
Again, only the initial five
entries are included in the sample Table 2 within this paper itself; the 
full
table is available electronically from the journal.

\subsection{X-ray Emitting AGN Having Narrower Permitted Emission Lines}

We have also examined the remainder of the 15,129 spectra of SDSS objects in 
RASS error circles, as some bona fide AGN identifications will fail to 
satisfy the broad-line (based on a simple single-component Gaussian 
pipeline fit) criterion discussed in the previous section.
Our experience from Paper 1 verified that examining all relevant spectra 
helps 
insure fuller inclusion of such
X-ray emitting AGN subclasses as NLS1s (see \S 4.1 for NLS1 details),
Sy~1.5s, 1.8s, and 1.9s,
i.e., objects with broad-line regions 
related to those of classic quasars and Seyfert 1s, but which may be 
observed to 
have ``narrower" ($<1000$ km~s$^{-1}$) permitted-line components as well. 
The catalog of ``narrow-line" AGN presented in this section is also
extended to
include Seyfert 2 and type 2 quasar candidates, whose optical 
spectra are even more strongly dominated by narrow emission lines.
The additional narrow line AGN cataloged in this section are mainly
from the lower redshift regime of our sample.

Following Paper 1, for classification among this narrower-lined group
as Sy~1.5 to Sy~1.8, we manually verify from each SDSS spectrum
that the full 
width near
the continuum level (hereafter, FWZI) of the H$\beta$ emission line 
exceeds 2500~km~s$^{-1}$, also verifying
that the width of the H$\beta$ line exceeds that of the 
[OIII] 5007 emission in the same object by at least 
$\sim$1000~km~s$^{-1}$. 
Hence 
each such object has some ``broad-line" component 
substantially wider 
than typical of its narrow line region. 

We also select among the narrow-line group, AGN 
whose SDSS 
optical spectra lack strong broad H$\beta$ emission, but 
which nonetheless 
do have a markedly broad (at least at the continuum level) 
H$\alpha$ emission component. As in Paper 1, we again refer to these 
objects loosely as Sy~1.9s, though many may be Sy 1.5-1.8, where 
S/N is merely low near H$\beta$. As in Paper 1, we again very 
conservatively limit our Sy~1.9 candidate list to include 
just those cases 
with very broad H$\alpha$, requiring 
FWZI(H$\alpha$)$>$6000~km~s$^{-1}$. 
(Note that many Sy 1.9s were properly identified 
by the DR5 pipeline as broad-lined AGN based on the FWHM of H$\alpha$, and 
so are 
already cataloged in the current paper under the broad-line criteria in \S 
3.1).

Our narrow-line catalog also includes
candidate X-ray emitting Sy~2s and related AGN. We reemphasize the
warning of Paper 1 that
historically a number of such possible cases found in earlier
X-ray surveys have, when  
subsequently scrutinized with improved optical spectroscopy,
ultimately been reclassified 
(e.g., Halpern, Turner, \& George 1999) into one of the categories 
already considered above, e.g., NLS1, Sy 1.8, or Sy 1.9. Inclusion
of these X-ray emitting Sy 2 candidates
may, at the very least however,
call attention to especially subtle cases of X-ray emitting  
Sy~1.8-1.9s or NLS1s that we otherwise might have missed.
For most Sy~2 candidates, we again
adopt the criteria of Kewley et al. 
(2001) based on the relative line strengths of [OIII]$\lambda5007/H\beta$, 
[NII]$\lambda6583/H\alpha$, and [SII]$\lambda6717,6731/H\alpha$; these
line ratios define regions in ``BPT diagrams"
(Baldwin et al. 1981) populated
by AGN versus starbursts. For a few objects, 
especially those at $z>0.5$
where $H\alpha$, [NII], and/or [SII] measures are not available,  
we require only that [OIII]$\lambda5007/H\beta > 3$. 
Note that we use the SDSS pipeline Gaussian
line flux measures in this application. 
Tables 3 and 4 respectively catalog empirical and derived information for 
the additional 515 X-ray emitting narrower lined AGN discussed in this 
section; 405 of these have
some observed broad-line component (e.g., NLS1s and Sy~1.5-1.9s), and an 
additional 110 are Sy 2 or 
type 2 quasar 
candidates. 
The comment column in Table 4 may include the
spectral subclass type;
these taxonomical classifications are tabulated
to clarify why these narrower line AGN
did not satisfy  the broad-line criteria discussed in section 3.1
Again, only sample tables are included within this paper, with the full 
tables available electronically. Figure 3 shows selected SDSS 
spectra from this group of identifications, reflecting 
some of the diversity among these
narrower lined X-ray emitting AGN.

\subsection{BL Lac Candidates}

We have also expanded our RASS/SDSS sample of BL~Lac candidates to
SDSS DR5. In Paper 1 we provided a list of 45 
X-ray emitting BL Lacs and candidates, and additional cases from
SDSS were presented as a subset 
of the more general spectroscopically-selected SDSS optical 
BL~Lac sample discussed in Collinge 
et al. (2005).
The rarity of BL Lacs demands large areal sky 
coverage such as that achievable from SDSS, and their unusual 
spectral 
energy distributions (with a lack of 
strong spectral features) generally allows efficient selection only
via multi-wavelength approaches---especially in radio, optical, 
and X-ray surveys  (e.g., Stocke et al. 1991, Perlman et al. 1996, 
Laurent-Muehleisen et al. 1997, Bade et al. 1998a; also see
review by Urry \& Padovani 1995).

We employ similar approaches within our `ROSAT\_A' 
target selection algorithm (as discussed in section~2), to obtain the 
updated DR5 sample of 
RASS/SDSS BL~Lac candidates discussed here. We find (or recover) 181 
objects we 
consider as 
{\it probable} X-ray emitting RASS/SDSS BL~Lacs. For these higher 
confidence cases: (1) the SDSS optical object is within $1'$ of a RASS 
source; (2)~the SDSS object is also a positional match to a radio source 
(conservatively taken as $<2''$ for matches to 
FIRST sources, or $<7''$ for matches to other radio catalogs); (3) our 
measures from the SDSS optical spectrum reveal no strong emission 
($EW<5$~\AA); and, (4)~either there is no CaII H\&K break/depression 
evident in 
the SDSS optical spectrum, or if present any such break must be
weak.  Following Stocke et al. (1991), we require the CaII H\&K break to 
have $C\le$~0.25, where 
$C=0.14 + 0.86(f_{\lambda,+}-f_{\lambda,-})/f_{\lambda,+}$~and 
where $f_{\lambda,-}$
and $f_{\lambda,+}$~are the 
average specific fluxes over the wavelength ranges 3750-3950 and 4050-4250 
\AA, respectively (e.g., see Landt et al. 2002, Dressler \& Schectman 
1987). We also find as {\it possible} BL~Lac candidates 
85 
additional objects that are within RASS error circles and which either: 
satisfy the first three criteria (X-ray/radio sources with no strong 
optical emission), but which have slightly larger CaII H\&K breaks with 
$0.25<C<0.4$ (see March\~{a} et al. 1996); or, have too low S/N in their 
SDSS optical spectra to claim with confidence the BL~Lac spectral nature 
of criteria (3)
and (4); or, which show approximately featureless 
SDSS spectra, i.e., that satisfy criteria (1), (3), and (4), but where 
there is no
close match to a radio source.
The latter seven objects are of potentially high interest
for additional follow-up as unusually weak radio sources; the vast
majority of
confidently-identified BL~Lacs cataloged thus far are radio sources
(though some are weak).

The 5740 deg$^2$ of sky considered here thereby include a total of 266 
candidate X-ray BL~Lac counterparts,
and basic information for them is provided in 
Tables 5 and 6. 
Again, only the initial 5 entries are included in these sample tables; the 
full tables are available electronically from the journal. An entry
of `zunc' in the 
comment column of Table 6 denotes cases where a
redshift from SDSS or the literature is unavailable or highly uncertain.
If a redshift is not obtained from the SDSS spectrum, nor 
available in the literature, we adopt $z=0.3$ (near the median of others 
in the sample) for estimating $\alpha_{ox}$ in Table 6; as this is 
essentially a distance-independent ratio of luminosities, the 
values of $\alpha_{ox}$ should be approximately correct in most cases,
except for precise values of spectral K-corrections. In 
Table 6, we denote the possible but less certain BL~Lac candidates as 
`BL?' in the comment column. Example SDSS spectra of representative 
X-ray BL~Lac 
candidates are shown in Figure 4.

\begin{figure}
\plotone{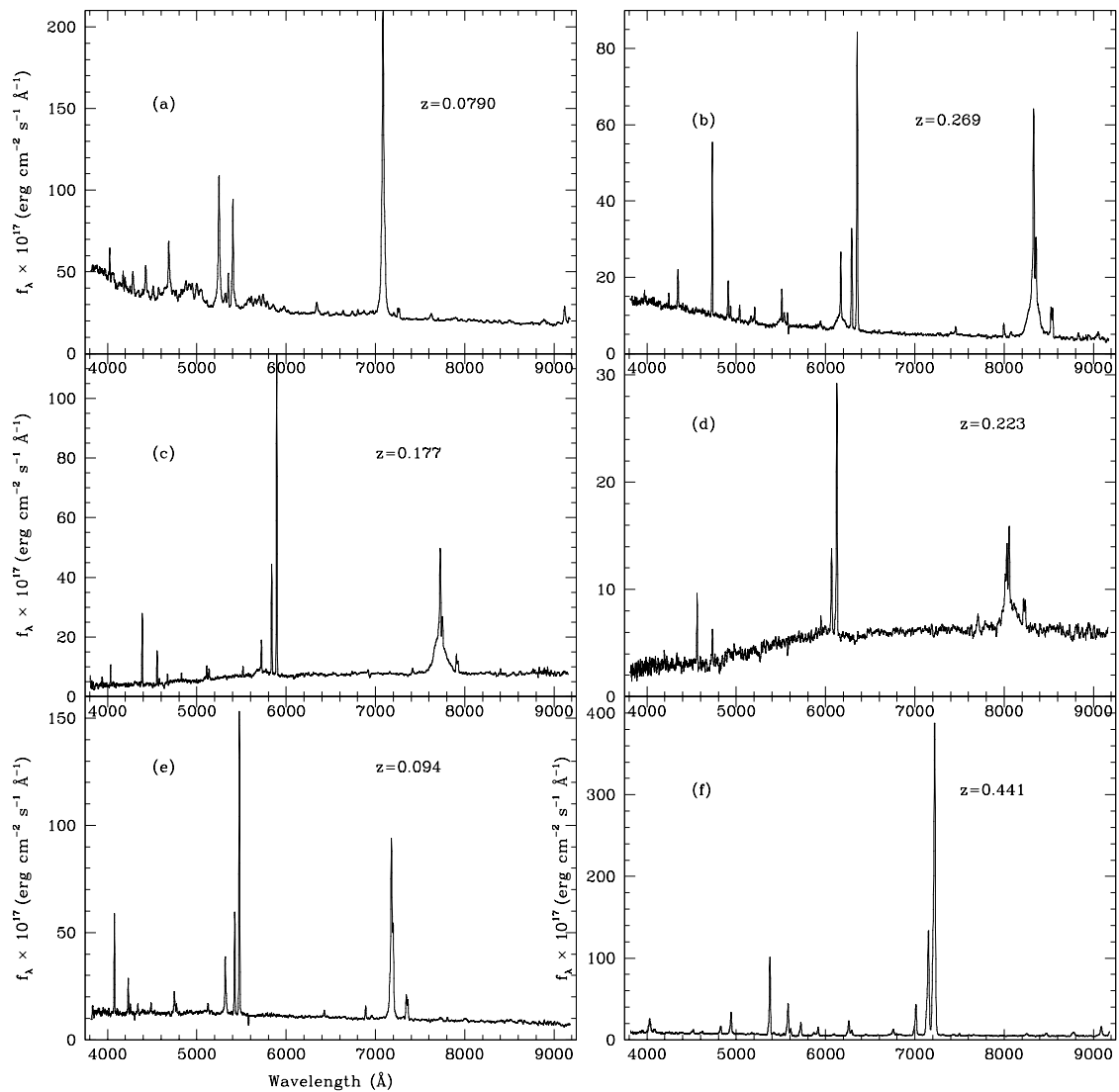}
\caption{
Selected SDSS optical spectra for
RASS/SDSS X-ray emitting quasars/AGN with narrower
permitted emission components, chosen to reflect the diversity
among the 515 objects cataloged in section 3.2.
Approximate spectral taxonomical classifications for (a-f) are,
respectively: NLS1, Sy 1.5, Sy 1.8, Sy 1.9, Sy 2, and a possible type~2
quasar (see Zakamska et al. 2003 for a discussion of SDSS type~2
quasars).
(a) SDSS J141755.54+431155.8; (b) SDSS J142337.63+341052.9;
(c)~SDSS J155021.40+295027.8; (d) SDSS J085348.18+065447.1;
(e) SDSS J143001.63+455049.1; (f) SDSS J091345.48+405628.2.
\label{fig3}}
\end{figure}

\begin{figure}
\plotone{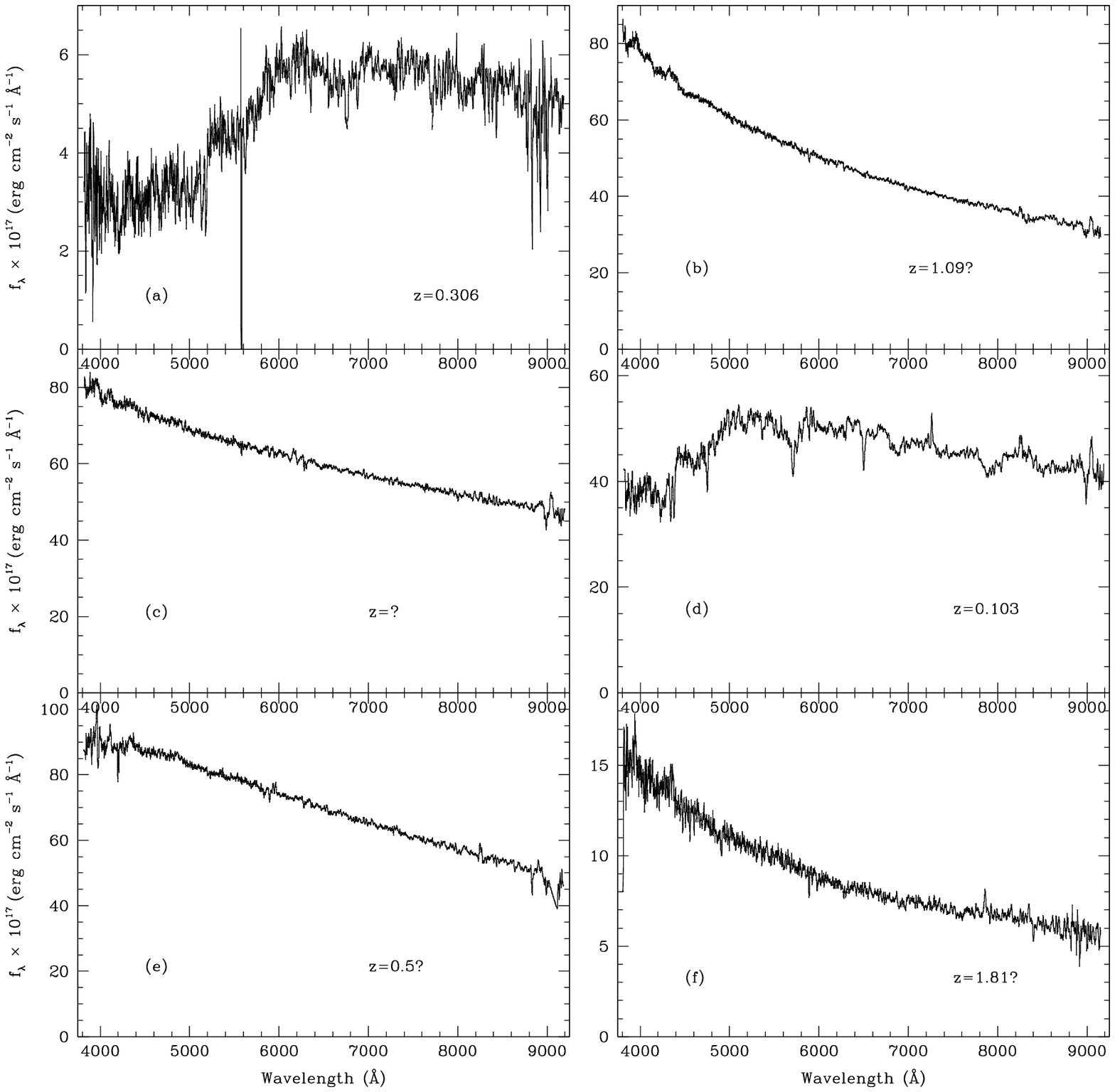}
\caption{
Representative SDSS optical spectra for
RASS/SDSS X-ray emitting BL~Lac candidates
as discussed in section 3.3.
Shown (a-e) are spectra for the first 5 objects listed in 
Tables~5-6, along with a possible high-redshift case (f).
A total of 266 such BL Lac candidates
are presented in our updated RASS/SDSS catalog.
(a) SDSS J002200.95+000657.9; (b) SDSS J003514.72+151504.1;
(c) SDSS J005041.31$-$092905.1; (d) SDSS J005620.07$-$093629.8;
(e) SDSS J014125.83$-$092843.6; (f) SDSS J124700.72+442318.7.
\label{fig4}}
\end{figure}

About half of these objects were first identified as BL~Lac candidates via 
SDSS, and/or have redshifts first reliably measured from SDSS spectroscopy. 
(Though some, for example, are cataloged in Bade et al. 1998b, 
but lacked secure identifications and/or redshifts in their 
low-resolution prism data). About 40\% of the full sample of 266 currently 
have reasonably confident SDSS or published spectroscopic redshifts, and 
another 20\% less confident spectroscopic redshift estimates.

As an added precaution to limit contamination of the BL Lac sample
by stars with weak features (e.g., DC white dwarfs),
we have also considered proper motion
information, as in Collinge et al. (2005).
We obtained proper motions from the SDSS database for 242 of our 266 
RASS/SDSS
BL~Lac candidates.  Of these 242,
only four BL Lac candidates appear to have significant proper motion 
($>$20~mas/yr).  However, three of the latter candidates have secure 
extragalactic redshifts from 
their SDSS spectra, and the fourth object is cataloged as a 
confirmed BL~Lac in Veron-Cetty \& Veron (2006). All
four candidates additionally match to both an X-ray and a radio source, 
suggesting little contamination by nearby hot white dwarfs.   Thus, 
it seems likely that most of the BL~Lac identifications are correct (see 
also section 5 below).

This catalog of RASS/SDSS/radio selected BL~Lac candidates constitutes one 
of the largest samples obtained to date (see also Collinge et al. 2005,
for a similarly large optically-selected sample from SDSS). 
Each is accompanied by
uniform X-ray, optical, and radio data.

\section{Other Rare Classes of X-ray Emitting AGN}

\subsection{Narrow-line Seyfert 1s (NLS1s)}

In our expanded DR5 RASS/SDSS sample 
we identify from SDSS spectra a total of 774 candidate X-ray 
emitting NLS1s. These include objects cataloged in both sections 3.1 and 
3.2, and they are denoted by `NLS1?' in the comment 
columns of Tables 2 and 4.
Williams et al. (2002)  discussed  45 of these SDSS X-ray NLS1s
independently selected from the SDSS Early Data Release, and we 
added 120 more in Paper~1. In the optical,
NLS1s have unusually narrow permitted lines, though in most
other ways resemble
Sy~1s more than Sy~2s; in the X-ray, they often show 
strong soft X-ray excesses and marked variability. Various
explanations have been suggested for NLS1s, e.g., unusually low-mass 
black holes and/or higher accretion rates relative to Eddington, etc. (see
reviews by Boller 2000 and Pogge 2000).

The 774 X-ray emitting candidate NLS1s all have the following spectral 
characteristics, representative of the NLS1 class (e.g., Pogge 
2000): [OIII]~$\lambda$5007 to $H\beta$ flux ratios of less 
than~1 (from the SDSS spectroscopic pipeline Gaussian-fit emission line 
measures), 
H$\beta$ FWHM (again from the SDSS spectroscopic pipeline) less than 
2000~km~s$^{-1}$, and strong optical Fe 
emission.
Seventy-four of these are detected in the FIRST radio survey; these are
of possible special interest, as radio-loud NLS1s may be unusually 
rare (e.g., Komossa et al. 2006).
An example SDSS spectrum for a NLS1 candidate is shown in Figure~3a.

\subsection{BALQSOs}
ROSAT studies extending back more than a decade (Green et al. 1995), along
with many subsequent investigations, demonstrated that BALQSOs
as a class are weak emitters in soft X-rays. The soft X-ray
deficiency is thought to arise
due to absorption in BAL material of high column
density, typically inferred equivalent to
$N_H \sim 10^{22-23}$~cm$^{-2}$ (e.g., Green et al. 2001;
Gallagher et al. 2002). In our DR5 sample,
there are 14 cases of traditional BALQSOs (e.g., those with 
``balnicity-index" $BI>0$) that fall within $1'$ RASS error circles, plus 
another two dozen cases with weaker BALs and/or possible mini-BALs.
(See Trump et al. 2006 for an extensive recent discussion and catalog of 
SDSS BALQSOs, selected based on both the standard Weymann et al. 1991 BI 
criteria, as well as the more inclusive ``absorption line" index or AI
criteria of Hall et al. 2002). Of 
course, in some cases an improved X-ray position may
be required to definitively establish whether or not all these 
BALQSOs are actually the X-ray source counterparts, given the expected
(see \S 5) few percent contamination of our full AGN sample.
Shown in Figure~5a is the SDSS spectrum 
of an example X-ray emitting BALQSO.

Although definitive conclusions await further confirmations of which 
BALQSOs are 
genuine X-ray sources and which are just chance coincidences, the 
RASS/SDSS sample size is now large enough to begin to provide some 
ensemble statistical tests of the (anticipated low) 
incidence of X-ray emitting BALQSOs. For example, the Trump et al. (2006) 
SDSS study found an incidence of 10.4\% (with a formal statistical 
error 
of about 0.2\%) for mainly optically-selected $BI>0$ BALQSOs among DR3 
quasars with $z>1.7$. The analogous BALQSO incidence in our RASS/SDSS 
X-ray selected sample is $4.6\% \pm1.5\%$, i.e., a significantly smaller 
incidence in a soft X-ray selected
sample (at least) at the $>3\sigma$ level. In the latter estimate we
consider just the DR3 subset for maximum 
consistency with the Trump et al. study; this is really an upper limit
on the incidence in the RASS/SDSS sample, as it somewhat optimistically 
assumes 
that 
all $BI>0$ cases are genuine 
X-ray/BALQSO associations.

\subsection{Unusual Line-Profile and Other Odd Cases}

There are many further quasars in our updated catalog with unusual optical 
line profiles (see Paper~1 for further discussion). 
These include further 
examples of X-ray emitting AGN with
weak/narrow H$\beta$ characteristic of Sy~1.8-2s but also having strong 
broad MgII (Figure 5b), 
post-starburst AGN (Figure 5c),
AGN with broad asymmetric Balmer profiles (Figures 5d-e), and
X-ray AGN with
highly unusual multiple-peaked emission line profiles (Figures 5e-f).

\begin{figure}
\plotone{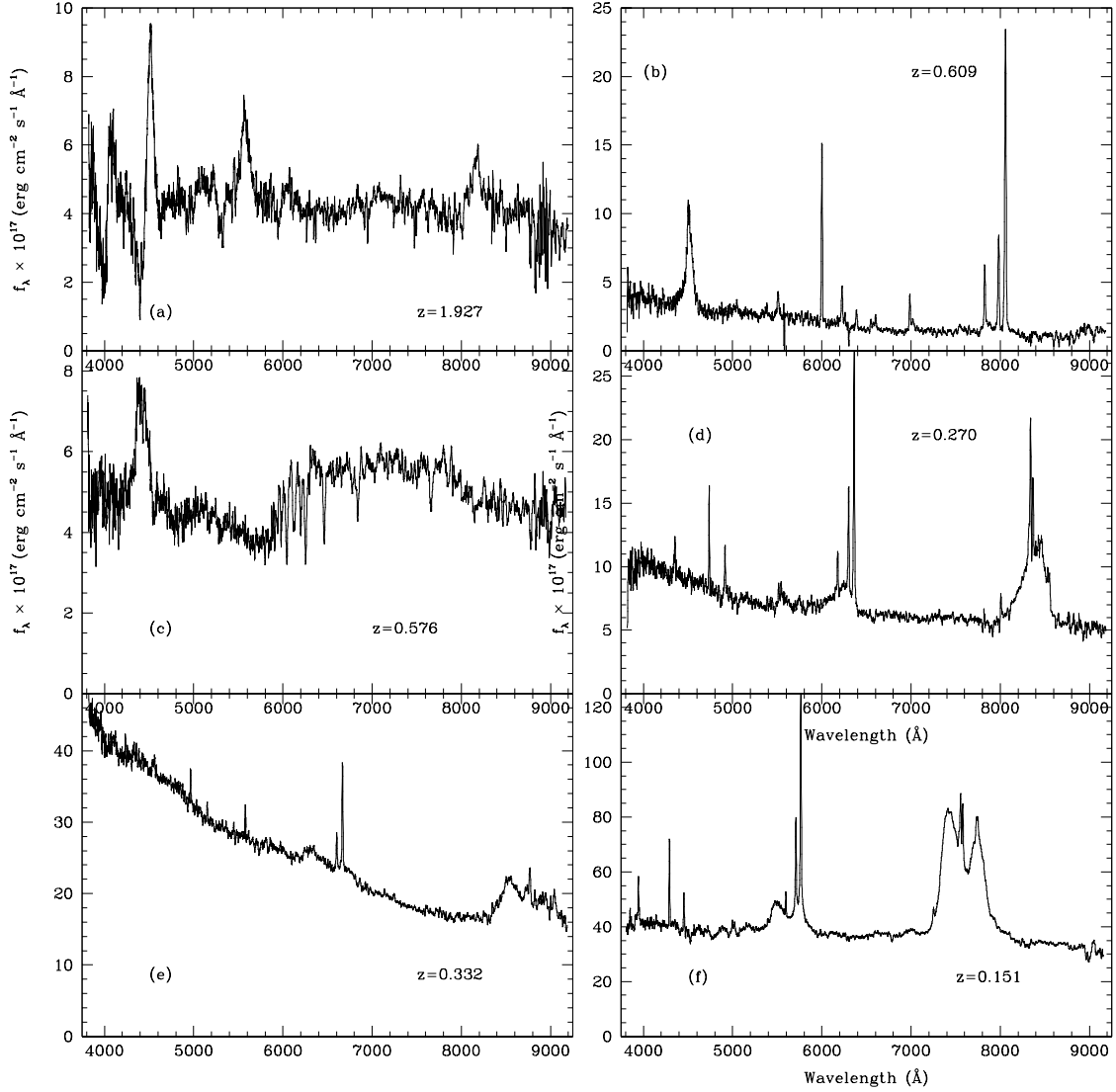}
\figcaption{
Selected other RASS/SDSS X-ray emitting AGN
having unusual
optical SDSS spectral characters. (a) SDSS J105626.95+482956.3 is example
of a possible X-ray emitting BALQSO. (b) SDSS J074951.42+204936.9 is an
AGN with weak/narrow H$\beta$ characteristic of Sy~1.8-2, but which has
strong, broad MgII. (c) SDSS J163446.49+461946.7 is a post-starburst AGN.
SDSS J115227.12+604817.4 (d) and SDSS J102738.53+605016.5 (e) are
examples of AGN with broad, asymmetric Balmer profiles. (f) SDSS
J161742.53+322234.3 is an example X-ray emitting AGN with unusual
multiple-peaked and broad emission lines profiles (SDSS J1027+6050 
may also be double-peaked).
\label{fig5}}
\end{figure}

\section{Ensemble Properties and Identification Reliability} 

From the expanded 5740 deg$^2$ of sky coverage in DR5,
7000 quasars or other AGN are
identified as likely RASS counterparts, each with uniform
and high-quality optical photometry and 
spectroscopy from SDSS, and X-ray data from ROSAT.
The ensemble X-ray flux, optical magnitude, and redshift 
distributions 
are depicted in Figure 6. The X-ray flux distribution (Figure 6a) 
reflects the typical depth of the RASS catalogs, with median
$f_x=2.5 \times 10^{-13}$~erg~s$^{-1}$~cm$^{-2}$; the optical
magnitude (Figure 6b) and redshift (Figure~6c) distributions
of the AGN remain typical of past 
identification work at comparable X-ray depth, with medians $g=18.8$ and 
$z=0.42$.
However, our updated RASS/SDSS catalog is sufficiently large to, for 
example, include 162 $z>2$ X-ray emitting AGN, 334
having $g<17$, and 505 with $f_x > 10^{-12}$~erg~s$^{-1}$~cm$^{-2}$
Note that in our sample, which is affected by both X-ray and optical flux
limits, luminosities and redshift are strongly coupled (Figure 7).

\begin{figure}
\plotone{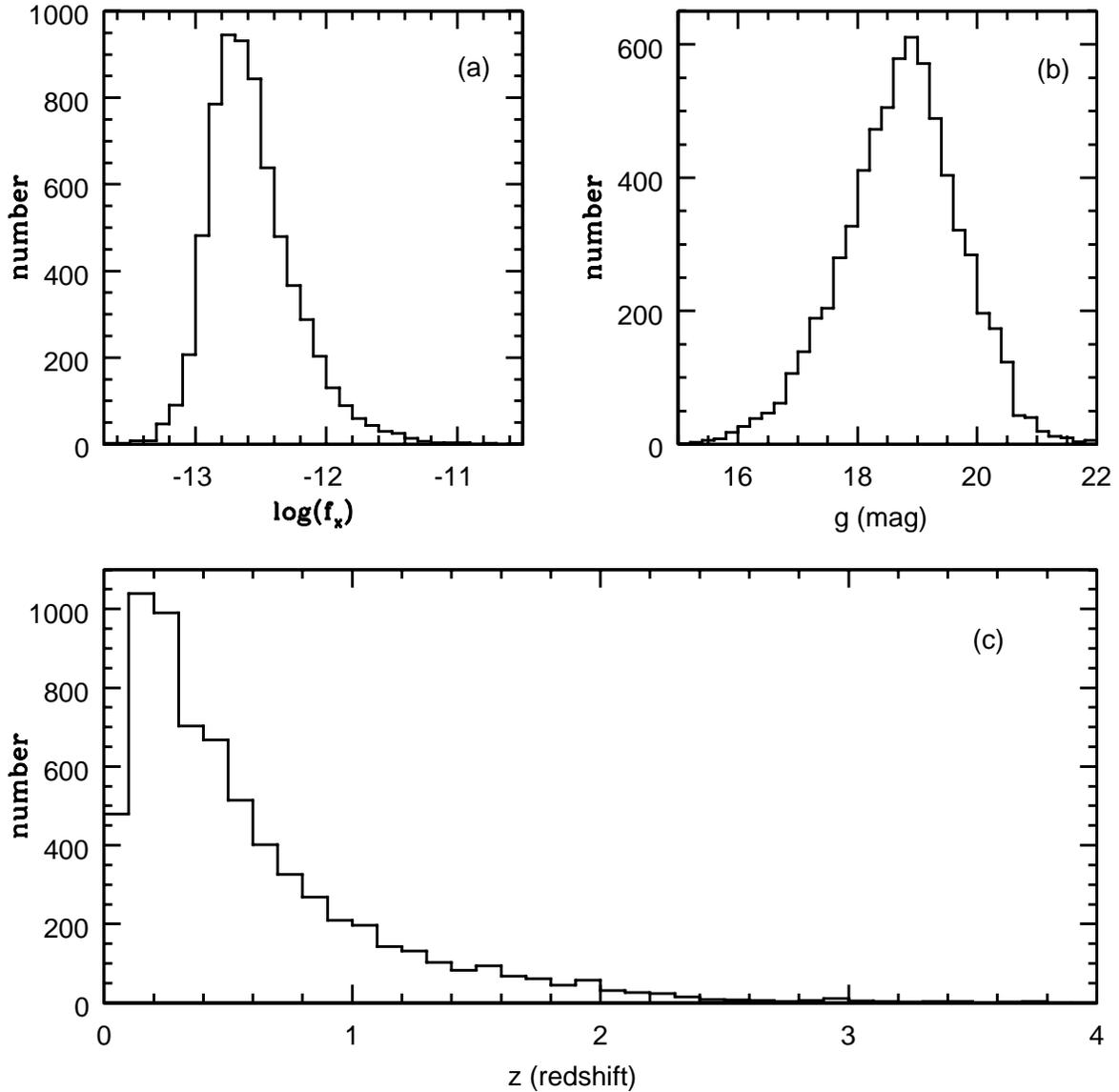}
\figcaption{
Distributions of 
(a) RASS X-ray fluxes, 
(b) SDSS $g$-band magnitudes,
and (c) redshifts for 6700 AGN counterparts of RASS/SDSS X-ray emitting
AGN. (BL Lacs are excluded from the plots here, as many have uncertain 
redshifts). The median magnitude ($g=18.8$) and redshift ($z=0.42$) are 
typical of other optical
identification efforts at comparable X-ray depth (median
$f_x=2.5 \times 10^{-13}$~erg~s$^{-1}$~cm$^{-2}$), though the
very large sample includes a substantial number of bright or
higher redshift AGN as well.
\label{fig6}}
\end{figure}

\begin{figure}
\plotone{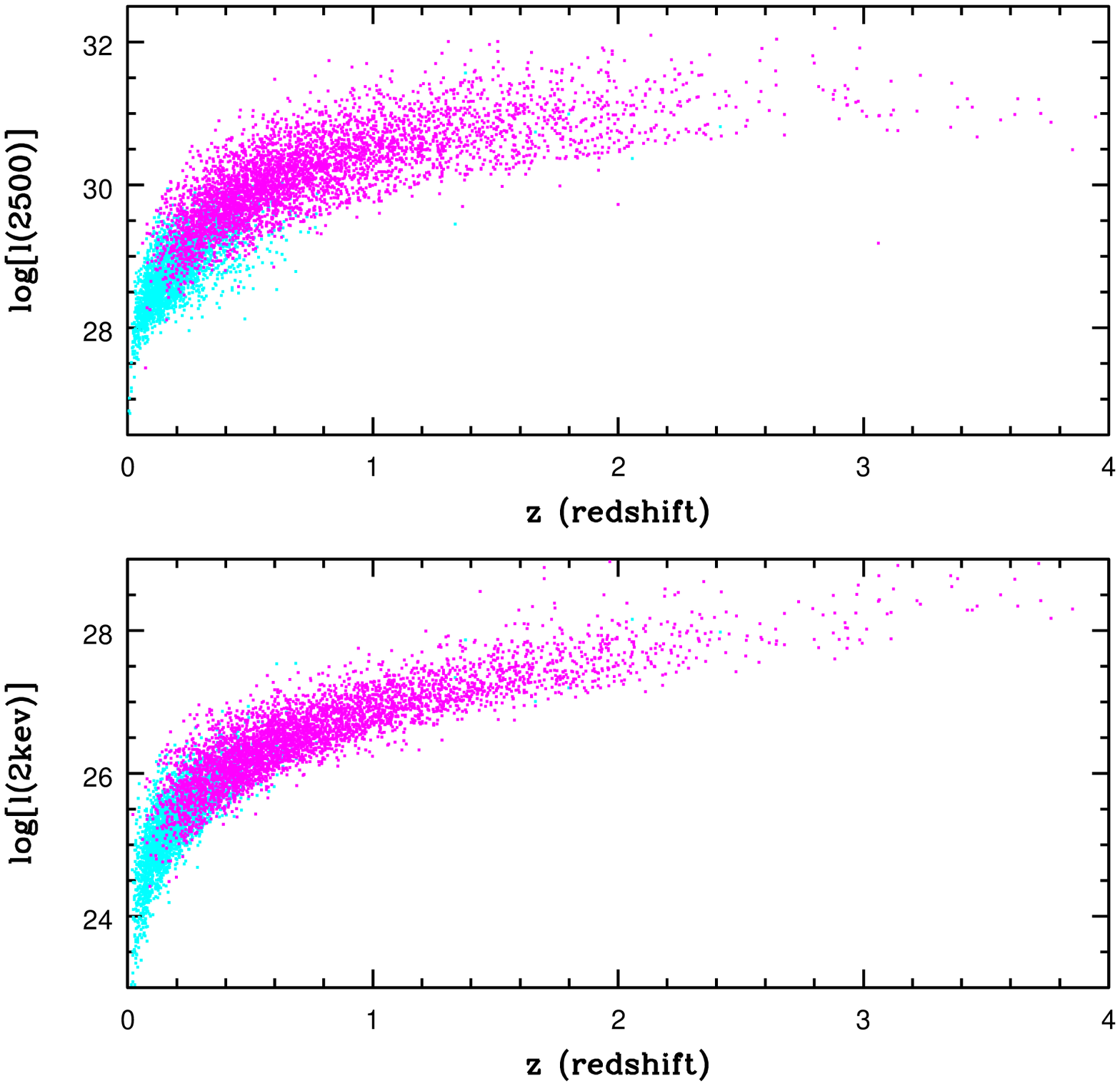}
\figcaption{
Luminosities and redshift are strongly coupled in our sample
with both X-ray and optical limits (cgs monochromatic luminosities
at frequencies corresponding to 2500 \AA\ and 2~keV are shown). 
The {\it magenta} points display data for
quasi-stellar X-ray identifications that are unresolved in SDSS
optical images, while {\it cyan} points show data for AGN 
that are morphologically
resolved in SDSS.
\label{fig7}}
\end{figure}

In our expanded DR5 catalog, the distributions of offsets between RASS 
X-ray positions 
and SDSS optical positions are again approximately as 
expected, if most of the 
7000 AGN are the proper identifications. For example, 86\% of the 
SDSS quasars/AGN fall within 30$''$ of the RASS X-ray positions (see 
the angular offset distribution in Figure 8a), in approximate agreement 
with the RASS 
positional uncertainty distribution,
independently derivable from Tycho stars also detected in RASS 
(Voges et al. 1999). In Figure~8b, these angular offsets have 
been normalized 
relative to their associated estimated RASS positional errors, to better
account for 
the 
dependence of the RASS positional uncertainty on the X-ray source 
significance, etc. This figure confirms that 
the RASS positional error estimates are reasonable, with 98\% of the 
suggested identifications at offsets smaller than 3$\sigma$ (where
$\sigma$ is the
estimated RASS X-ray positional error); we also note, however, that the 
distribution of RASS positional errors may not be fully
characterized by either a simple 1- or 2-dimensional Gaussian 
(Ag\"ueros et al. 2006). 
Figure 9 shows the
distributions of the squares, $r^2$, of the offsets 
between the SDSS optical positions and the RASS X-ray source 
positions for various AGN subclasses considered herein;
specifically, we display the 
fraction of suggested SDSS counterparts
falling within equal area annuli offset from the RASS X-ray 
source positions.
For a chance superposition of SDSS objects within
RASS error circles, these histograms would be approximately flat
with $r^2$. But the respective distributions for broad-line AGN (from
\S 3.1), narrow line AGN (\S 3.2), BL Lacs (\S 3.3), and NLS1s (\S 4.1)
are each strongly-peaked at small 
$r^2$ values, as expected if these AGN are statistically the proper 
X-ray source identifications.

\begin{figure}
\plotone{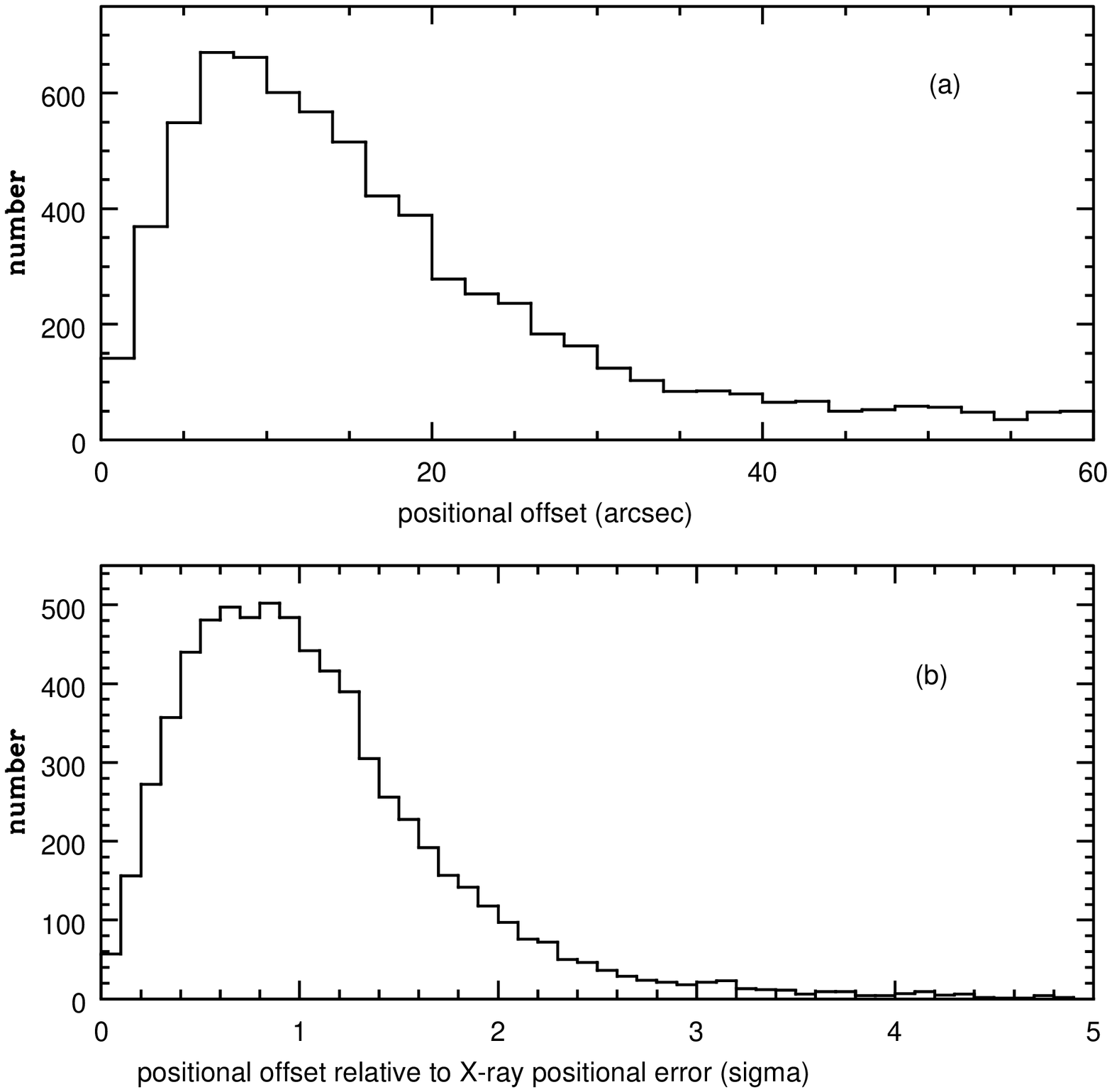}
\figcaption{
The distribution of differences between the
SDSS optical and RASS X-ray positions of the 7000 AGN
cataloged here are consistent with expectations,
if most are the proper
identifications. (a) The positional offset distribution in arcseconds
is approximately as expected for the RASS positional accuracy.
(b)~The distribution
of the differences between the
SDSS optical and RASS X-ray positions,
here normalized by their associated RASS X-ray source positional errors.
(A handful of objects with relative offsets $>5\sigma$ are excluded
from this plot).
\label{fig8}}
\end{figure}

\begin{figure}
\plotone{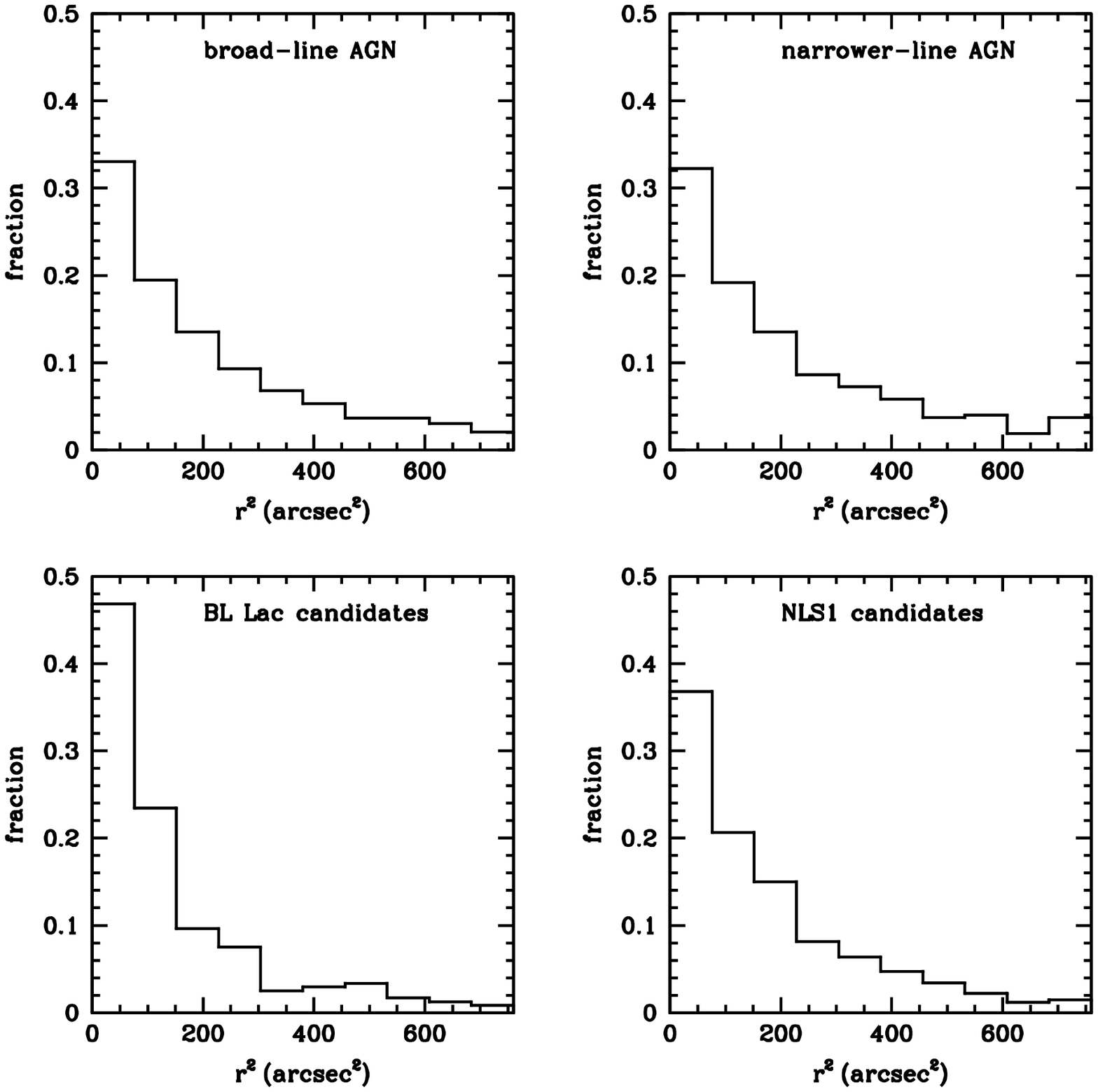}
\figcaption{
Distributions of the squares, $r^2$, of the
offsets between the SDSS optical positions and the RASS X-ray source 
positions;
specifically, we display the fraction of optical counterparts falling 
within equal 
area
annuli offset from the RASS X-ray source positions.
(We consider only objects within $r<27.5''$,
where all relevant algorithms for SDSS spectroscopy may select targets).
Distributions are
shown for quasars/AGN with predominant broad-lines discussed
in section 3.1 (upper left), quasars/AGN with narrower permitted
lines discussed in section 3.2 (upper right), BL~Lac
candidates discussed in section 3.3 (lower left), and candidate NLS1s
discussed in section 4.1 (lower right). In each
case, the
distributions are very strongly-peaked at small $r^2$ values,
as expected if these AGN are statistically the proper
X-ray source identifications.
\label{fig9}}
\end{figure}

Further quantifying the statistical reliability 
of these identifications is the following. The surface density of SDSS 
optically-selected quasars is a little more than 12~deg$^{-2}$ (e.g., 
Schneider et al. 2005), while the combined area covered by all RASS error 
circles considered in the 5740~deg$^2$ area of this updated sample is a 
little over 13~deg$^2$. Thus, based on surface density arguments, only a 
small fraction (of order 3\%) of the proposed quasar/AGN X-ray source 
counterparts are likely to be spurious random chance positional 
coincidences. Similarly, if we randomly alter the positions in the RASS 
catalog by several arcminutes, and then cross-correlate again between 
SDSS quasars and such synthetically-offset RASS catalogs, we 
alternately estimate that 
only of order 5\% of the SDSS AGN cataloged herein are likely to be 
chance positional 
coincidences, unrelated to RASS X-ray sources.

The ratios, $f_{x}/f_{opt}$, of X-ray to optical flux for the 
identifications are also as expected for typical X-ray emitting AGN. 
Roughly as much energy is emitted in the X-ray as in the optical bands for 
typical quasars, as is reaffirmed by the empirical $f_{x}/f_{opt}$ 
distribution shown in Figure~10a for the 6224 quasars/AGN with predominant 
broad emission line regions considered in section 3.1 In estimating 
$f_{x}/f_{opt}$, we here adopt the (corrected) 0.1-2.4~keV X-ray flux 
(e.g., from Table 2), and estimate the optical broadband flux in a 
4000-9000 \AA\ bandpass using the $g$-band PSF magnitudes (e.g., again 
from Table 2) assuming an optical power law in energy with index 
$\alpha_o=0.5$. The observed $f_{x}/f_{opt}$ distribution in Figure 10a 
for DR5 broad-line RASS/SDSS AGN appears very similar to that found for 
EMSS quasars (see Stocke et al. 1991); two-sided Kolmogorov-Smirnov (K-S) 
comparisons between the EMSS and RASS/SDSS $f_{x}/f_{opt}$ distributions 
for quasars confirm similarity at the 23 to 61\% level, for a plausible 
range of conversions between the differing EMSS versus RASS/SDSS X-ray 
and 
optical passbands (and implicit spectral assumptions). Figure~10b shows 
the analogous distribution of $f_{x}/f_{opt}$ ratios for the 266 RASS/SDSS 
BL~Lac candidates discussed in section 3.3. This distribution also appears 
very similar to that found in earlier X-ray selected BL~Lac samples (e.g., 
again see Stocke et al. 1991); two-sided K-S tests indicate a match of 
EMSS 
and RASS/SDSS $f_{x}/f_{opt}$ distributions for BL~Lacs at the 21 to 64\% 
level, with the range again reflecting uncertainties in conversions 
between systematically different passbands, etc. The broad agreement of 
the $f_{x}/f_{opt}$ distributions with previous work for both quasars and 
BL~Lacs further confirms that the vast majority of the suggested RASS/SDSS 
identifications are likely correct.

\begin{figure}
\plotone{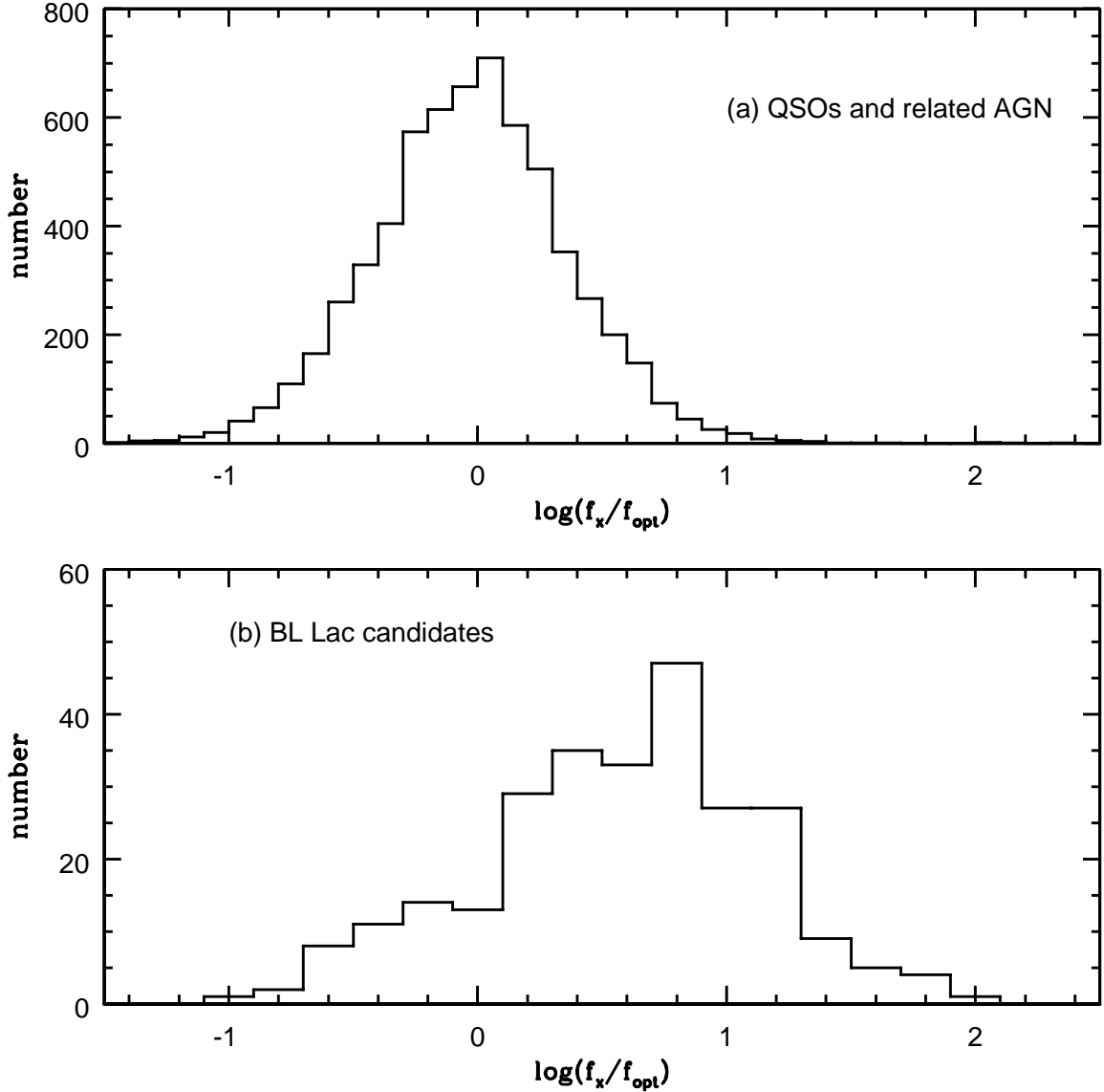}
\figcaption{
 (a) The $f_{x}/f_{opt}$ distribution
for the 6224 quasar/AGN identifications having predominant
broad
emission line regions (see \S 3.1).
The RASS/SDSS quasars/AGN emit approximately as much energy
in the X-ray as in the optical band,
as expected if these are the proper identifications.
(b)~The $f_{x}/f_{opt}$ distribution for the
RASS/SDSS BL~Lac candidates discussed in section 3.3. The
distribution is
similar to that found in other X-ray selected BL~Lac surveys,
affirming
that these RASS/SDSS objects are also likely to be the proper 
identifications.
\label{fig10}}
\end{figure}

\section{An Example X-ray/Optical Correlation from the Expanded Sample}

This paper primarily presents  
updated catalog information, but we also include below
a brief update on our Paper 1 
example of an X-ray/optical correlation. More generally, the large 
RASS/SDSS sample size and associated
uniformity of X-ray and optical (and radio) data may be useful for 
other more detailed, multiwaveband follow-on studies (e.g., see K\"ording 
et al. 2006).

The 6224 X-ray emitting AGN with 
predominant broad-line regions discussed in section 
3.1 show (see Figure 11)
the well-known correlation 
between (the logarithms of) optical and X-ray monochromatic 
luminosities found in many 
earlier studies/samples
(e.g., Avni \& Tananbaum 1986). We again caution (see 
Paper 1)
that our sample is, of course, an
X-ray selected sample with inherent X-ray biases. Nonetheless,
the best-fit linear relationship we find (Figure 11a)
between the logarithms of X-ray and optical luminosities,
equivalent to 
$l_{x} \propto l_{opt}^{0.90\pm0.01}$, is in good
agreement with that found in other studies 
(e.g., see Wilkes et al. 1994; Green et al. 1996; Vignali et al. 2003; 
Paper 1). This
best-fit linear relation is 
virtually unchanged 
when excluding the 15\% of the objects that are radio-detected.

\begin{figure}
\plotone{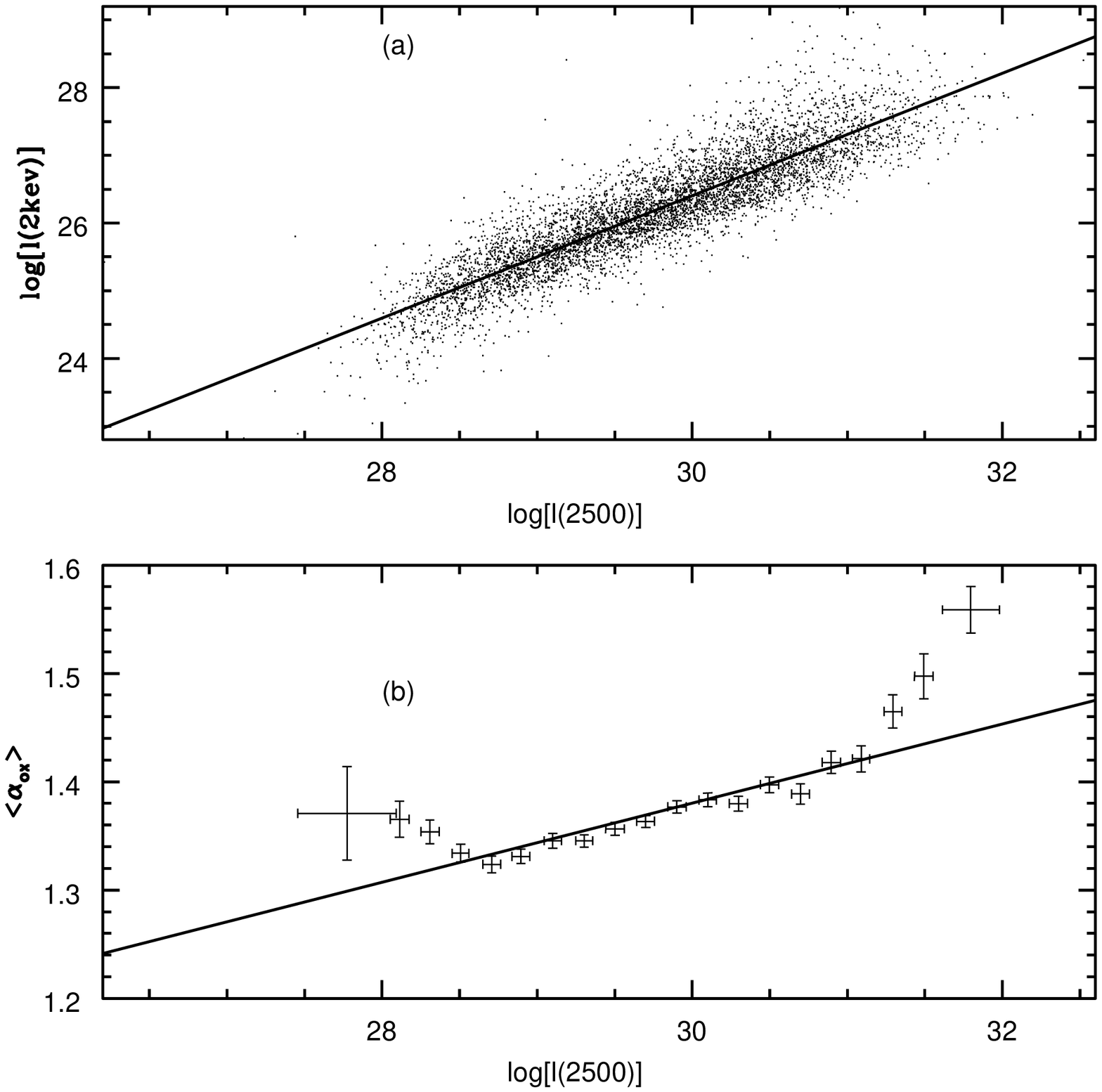}
\figcaption{\small{
The well-known relationship between X-ray and optical wavebands 
(logarithms of monochromatic luminosities in cgs units at 2~kev and 
2500\AA\ are used) is also seen here in our X-ray selected sample 
of 6224 quasars/AGN with predominant broad emission line regions. 
(a)~The solid line is a linear least-squares fit to the logarithms 
of $l_{2kev}$ versus $l_{2500}$ data, with slope $0.90\pm0.01$. 
(b)~Consistent results are obtained when regressing $\alpha_{ox}$ 
against optical luminosity, and in this case a simple linear 
regression yields $\alpha_{ox} \propto l_{opt}^{0.037}$, equivalent 
as expected to $l_{x} \propto l_{opt}^{0.90}$. The solid line is 
the best-fit linear regression relation when fit to all 6224 points 
separately; the error bars show the mean and the standard error in 
the mean value of $\langle \alpha_{ox} \rangle$, as well as the mean and 
standard 
deviations in optical luminosity, when considering averages taken 
in various logarithmic optical luminosity bins (typically with 100 
to 500 points per bin). The deviation of multiple adjacent, and 
independent, error bars significantly away from the simple best-fit 
line suggests the possibility of a more complicated relation than 
assumed in standard linear regression models.
\label{fig11}}}
\end{figure}

As we suggested previously in Paper 1, however, the best-fit slopes of the
relation within our RASS/SDSS 
sample differ
when separately considering lower versus higher optical-luminosity 
objects. For example, if we divide the 6224 broad line AGN (\S 3.1) into 
two subsamples at the median
value $log(l_{opt})=29.76$ (cgs),
we find $l_{x} \propto l_{opt}^{0.96\pm0.01}$
for the half below the median, but $l_{x} \propto l_{opt}^{0.85\pm0.02}$
for the half above the median optical luminosity;
both these slope values are consistent with those
we quoted in Paper 1---though from a much smaller sample in that
earlier paper---for a 
very similar 
luminosity division. If one restricts
consideration to just the 406 objects in our DR5 sample with highest 
optical luminosity $log(l_{opt})>31$, the 
best-fit slope of the logarithmic relation
decreases even further, corresponding to $l_{x} \propto 
l_{opt}^{0.50\pm0.09}$. Very 
recently, some large studies (Steffen et al. 2006)
based on optically-selected samples have also begun
to confirm that the best-fit slope in the relationship between 
$log(l_{x})$ and $log(l_{opt})$
may not be a constant, but
instead may itself depend on optical luminosity.
Such a 
``non-linear" relationship between $log(l_{x})$ and $log(l_{opt})$
was suggested in our Paper 1
and earlier studies such as those of Yuan et al. (1998a). Such
complicated possible dependences caution that when 
considering results among 
different 
studies, it is useful to compare X-ray versus optical correlations for
objects of similar luminosity.

One might alternately divide the sample between 
morphologically-resolved 
AGN and those that have optical stellar-PSF morphology in SDSS images (see 
the type parameter in
Table 1). With this sub-division, one finds a similar result:
$l_{x} \propto l_{opt}^{0.97\pm0.02}$ for resolved AGN compared with 
$l_{x} \propto l_{opt}^{0.84\pm0.01}$ for stellar-morphology AGN.
Not too surprisingly, however, the resolved AGN (for which optical host 
galaxy photometric contributions may be most problematic) are at 
systematically 
lower optical luminosity and redshift than the optically-unresolved 
subset (Figure 7).
Further studies will be required to disentangle such strongly 
coupled parameters.

Regressing  $\alpha_{ox}$ 
against optical luminosity yields similar results. For the
6224 quasars/AGN with observed broad-line regions
discussed in section 3.1,
this regression yields  $\alpha_{ox} \propto l_{opt}^{0.037\pm0.002}$, 
approximately equivalent as expected to $l_{x} \propto l_{opt}^{0.90}$.
Figure 11b depicts the relationship from the $\alpha_{ox}$ perspective.
The solid line shows the above linear best-fit regression 
relation with slope 0.037,
while the error bars show the mean and the standard error
in the
mean value of $\alpha_{ox}$, as well as the mean and standard deviations 
of the optical
luminosity, when considering averages taken in various 
logarithmic optical luminosity
bins. (The vertical error bars account for the
large number of data points averaged to estimate $\langle \alpha_{ox} 
\rangle$ in each 
bin: typically 100-500, with a range of 42 to 534).
The offset of multiple independent and adjacent error bars significantly 
away from the simple best-fit 
line in Figure 11b, further suggests the possibility of a more complicated 
relation 
than assumed in the standard linear regression of $\alpha_{ox}$ versus 
log($l_{opt}$),
or in the equivalent log($l_x$) versus log($l_{opt}$) relation discussed
above.

However, we reiterate our caution of Paper 1
that the multiple strongly correlated
parameters involved here require a much more careful 
analysis, including proper accounting and disentangling of
the various selection biases, 
intertwined dependences on redshift and optical
luminosity, disparate dispersions
in optical compared with X-ray luminosities 
(e.g., Yuan, Siebert, \& Brinkmann 1998b), host galaxy 
photometric contamination, radio dependences, and so on. To 
highlight such 
complications,
we note that two recent studies on optically-selected SDSS quasars
having ROSAT information available arrive at rather different 
conclusions about whether $\alpha_{ox}$ may also depend 
on redshift as well as optical luminosity (Shen et al. 2006; Steffen 
et al. 2006). As noted
above, certainly these two parameters (and others) are
strongly correlated in our RASS/SDSS sample, with flux limits in
both the X-ray and the optical (Figure 7).

\section{Summary}

The SDSS optical and RASS X-ray surveys are well-matched
to each other, allowing
efficient large-scale identification of X-ray source
optical counterparts. 
Application of our approach to
SDSS DR5 data has provided
homogeneous identification and RASS/SDSS flux and
spectroscopic data for a large sample of 
X-ray emitting quasars and other kinds of AGN. 
The combination of SDSS multicolor selection
and RASS data---and in some cases FIRST radio information---is 
highly 
efficient for the selection of X-ray emitting quasars/AGN.
In our updated analysis encompassing 5740~deg$^2$ of sky, 
7000 plausible X-ray emitting quasars/AGN have been
optically identified, including hundreds of rare cases
such as BL~Lacs and NLS1s. The RASS/SDSS
survey is rapidly approaching 
$\sim10^4$ fully and homogeneously characterized 
optical counterpart identifications. 
The large sample will allow for a variety of more detailed
studies of various AGN subclasses and individual objects of special
interest,
as well as for studies of ensemble correlations between optical and X-ray
wavebands. In closing, we also note that many of the cataloged 
objects have X-ray fluxes accessible to at least the next generation 
of envisioned high-quality X-ray spectroscopy experiments, such as 
Constellation-X.

\bigskip
\acknowledgments
\noindent{\it Acknowledgments:} 
S.F. Anderson and R.M. Plotkin gratefully acknowledge support
from NASA/ADP grant NNG05GC45G. We thank Jonathon Trump for useful
discussions.

Funding for the SDSS and SDSS-II 
has been provided by the Alfred P. 
Sloan Foundation, the Participating Institutions, the National Science 
Foundation, the U.S. Department of Energy, the National Aeronautics and 
Space Administration, the Japanese Monbukagakusho, the Max Planck Society, 
and the Higher Education Funding Council for England. The SDSS Web Site is 
http://www.sdss.org/.

    The SDSS is managed by the Astrophysical Research Consortium for the 
Participating Institutions. The Participating Institutions are the 
American Museum of Natural History, Astrophysical Institute Potsdam, 
University of Basel, Cambridge University, Case Western Reserve 
University, University of Chicago, Drexel University, Fermilab, the 
Institute for Advanced Study, the Japan Participation Group, Johns Hopkins 
University, the Joint Institute for Nuclear Astrophysics, the Kavli 
Institute for Particle Astrophysics and Cosmology, the Korean Scientist 
Group, the Chinese Academy of Sciences (LAMOST), Los Alamos National 
Laboratory, the Max-Planck-Institute for Astronomy (MPIA), the 
Max-Planck-Institute for Astrophysics (MPA), New Mexico State University, 
Ohio State University, University of Pittsburgh, University of Portsmouth, 
Princeton University, the United States Naval Observatory, and the 
University of Washington.


\clearpage

\thispagestyle{empty}

\setlength{\voffset}{2cm}

\rotate
\begin{table}
\setlength{\tabcolsep}{0.06in}
\tablenum{1}
\caption{Observed Parameters of Broad-Line RASS/SDSS AGN\tablenotemark{a}}
\begin{tabular}{llrrrrrcrrrrrrr}
\tableline
\tableline
 RASS X-ray & SDSS optical & $u$ & $g$ & $r$ & $i$ & $z$ & opt & red- & X-ray & X & X & X & X & $f_x\times$ \\
 source & counterpart & & & & & & morph & shift & count & exp & HR1 & HR2 & like & $10^{13}$ \\
 RXS J & SDSS J& & & & & & & & rate & tim & & & & \\
 (1) & (2) & (3) & (4) & (5) & (6) & (7) & (8) & (9) & (10) & (11) & (12) & (13) & (14) & (15)\\
\tableline
000011.9+000223 & 000011.96+000225.3 & 18.24 & 17.97  & 18.02 & 17.96 & 17.91 & 6 & 0.479 & 0.0219  & 364 & -0.01 & -1.00 & 9 & 2.20 \\
000024.1+152026 & 000024.02+152005.4 & 19.41 & 19.18 & 18.99 & 19.08 & 19.13 & 6 & 0.989 & 0.0133 & 463 & -0.05 & -1.00 & 8 & 1.41 \\
000100.8$-$102318 & 000102.18$-$102326.9 & 18.98 & 18.69 & 18.33 & 18.24  & 17.66 & 6 & 0.294 & 0.0284 & 340 & 0.37 & -0.13 & 8 & 2.76 \\
000117.2+141150 & 000116.00+141123.0 & 19.01 & 18.68 & 18.70 & 18.56 & 18.16 & 6 & 0.404 & 0.0155 & 405 & 1.00 & 0.14 & 8 & 1.74 \\
000133.1+145601 & 000132.83+145608.0 & 19.22 & 19.20 & 19.25 & 19.04 & 18.39 & 3 & 0.399 & 0.0388 & 473 & 0.18 & -0.86 & 21 & 4.28 \\
\tableline
\end{tabular}
\tablenotetext{a}{[The complete version of this table is in the 
electronic
edition of the journal. The printed version contains only a sample.]}
\end{table}

\clearpage

\setlength{\voffset}{0cm}

\begin{table}
\setlength{\tabcolsep}{0.06in}
\tablenum{2}
\caption{Derived Parameters of Broad-Line RASS/SDSS AGN\tablenotemark{a}}
\begin{tabular}{llrrrrrrrl}
\tableline
\tableline
 RASS X-ray & SDSS optical & $g_o$ & red- & $f_x^c \times$ & $log(L_x)$ & $log(l_{opt})$ & $log(l_{x})$ & $\alpha_{ox}$ & comment \\
 source &  counterpart & & shift & $10^{13}$ & & 2500\AA\ & 2~kev &  & \\
 RXS J & SDSS J & &  & & & & & \\
 (1) & (2) & (3) & (4) & (5) & (6) & (7) & (8) & (9) & (10)\\
\tableline
000011.9+000223 & 000011.96+000225.3 & 17.85 & 0.479 & 5.42 & 44.76 & 30.14 & 26.22 & 1.50 & LBQS 2357-0014\\
000024.1+152026 & 000024.02+152005.4 & 19.02 & 0.989 & 3.60 & 45.41 & 30.37 & 26.88 & 1.34 & ...\\
000100.8$-$102318 & 000102.18$-$102326.9 & 18.53 & 0.294 & 6.60 & 44.32 & 29.40 & 25.78 & 1.39 & ... \\
000117.2+141150 & 000116.00+141123.0 & 18.46 & 0.404 & 4.68 & 44.51 & 29.73 & 25.97 & 1.44 & ...\\
000133.1+145601 & 000132.83+145608.0 & 19.03 & 0.399 & 11.36 & 44.88 & 29.49 & 26.34 & 1.21 & ...\\
\tableline
\end{tabular}
\tablenotetext{a}{[The complete version of this table is in the 
electronic
edition of the journal. The printed version contains only a sample.]}
\end{table}

\clearpage

\thispagestyle{empty}

\setlength{\voffset}{2cm}

\begin{table}
\setlength{\tabcolsep}{0.06in}
\tablenum{3}
\caption{Observed Parameters of RASS/SDSS AGN Having Narrower
Permitted Emission\tablenotemark{a}}
\begin{tabular}{llrrrrrcrrrrrrr}
\tableline
\tableline
 RASS X-ray & SDSS optical & $u$ & $g$ & $r$ & $i$ & $z$ & opt & red- & X-ray & X & X & X & X & $f_x\times$ \\
 source & counterpart & & & & & & morph & shift & count & exp & HR1 & HR2 & like & $10^{13}$ \\
 RXS J & SDSS J& & & & & & & & rate & tim & & & & \\
 (1) & (2) & (3) & (4) & (5) & (6) & (7) & (8) & (9) & (10) & (11) & (12) & (13) & (14) & (15)\\
\tableline
000202.5$-$103030 & 000202.95$-$103038.0 & 18.06 & 17.59 & 17.25 & 16.81 & 16.63 & 3 & 0.103 & 0.0610 & 324 & -0.12 & -0.14 & 31 & 5.94 \\
000250.8+000824 & 000251.60+000800.7 & 20.54 & 19.48 & 18.67 & 18.21 & 17.88 & 3 & 0.107 & 0.0410 & 388 & -0.32 & 0.09 & 9 & 4.10 \\
001056.7$-$090100 & 001056.25$-$090109.9 & 19.14 & 18.06 & 17.57 & 17.10 & 16.90 & 3 & 0.081 & 0.0283 & 343 & 0.49 & 1.00 & 10 & 2.82 \\
001618.5+011528 & 001617.83+011522.0 & 21.06 & 19.67 & 18.69 & 18.18 & 17.88 & 3 & 0.217 & 0.0398 & 402 & -0.13 & -0.14 & 18 & 3.78 \\
002608.2$-$000544 & 002608.38$-$000547.0 & 20.73 & 19.57 & 18.81 & 18.32 & 17.93 &  3 & 0.107 & 0.0228 & 415 & 0.81 & 0.49 & 13 & 2.11 \\

\tableline
\end{tabular}
\tablenotetext{a}{[The complete version of this table is in the 
electronic
edition of the journal. The printed version contains only a sample.]}
\end{table}

\clearpage

\setlength{\voffset}{0cm}

\begin{table}
\setlength{\tabcolsep}{0.06in}
\tablenum{4}
\caption{Derived Parameters of RASS/SDSS AGN Having Narrower
Permitted Emission\tablenotemark{a}}
\begin{tabular}{llrrrrrrrl}
\tableline
\tableline
 RASS X-ray & SDSS optical & $g_o$ & red- & $f_x^c \times$ & $log(L_x)$ & $log(l_{opt})$ & $log(l_{x})$ & $\alpha_{ox}$ & comment \\
 source &  counterpart & & shift & $10^{13}$ & & 2500\AA\ & 2~kev &  & \\
 RXS J & SDSS J & &  & & & & & \\
 (1) & (2) & (3) & (4) & (5) & (6) & (7) & (8) & (9) & (10)\\
\tableline
000202.5$-$103030 & 000202.95$-$103038.0 & 17.42 & 0.103  & 14.25 & 43.61 & 28.87 & 25.07 & 1.46 & Sy1.8,radio\\
000250.8+000824 & 000251.60+000800.7 & 19.35 & 0.107 & 10.03 & 43.49 & 28.13 & 24.96 & 1.22 & Sy1.5\\
001056.7$-$090100 & 001056.25$-$090109.9 & 17.92 & 0.081 &  6.87 & 43.07 & 28.45 & 24.53 & 1.50 & Sy1.8\\
001618.5+011528 & 001617.83+011522.0 & 19.56 & 0.217 &  8.89 &  44.13 & 28.70 & 25.60 & 1.19 & Sy1.8\\
002608.2$-$000544 & 002608.38$-$000547.0 & 19.48 & 0.107 & 4.87 & 43.18 & 28.08 & 24.64 & 1.32 & Sy2?,radio\\

\tableline
\end{tabular}
\tablenotetext{a}{[The complete version of this table is in the 
electronic
edition of the journal. The printed version contains only a sample.]}
\end{table}

\clearpage

\pagestyle{empty}

\setlength{\voffset}{2cm}

\begin{table}
\setlength{\tabcolsep}{0.06in}
\tablenum{5}
\caption{Observed Parameters of RASS/SDSS BL~Lac Candidates\tablenotemark{a}}
\begin{tabular}{llrrrrrcrrrrrrr}
\tableline
\tableline
 RASS X-ray & SDSS optical & $u$ & $g$ & $r$ & $i$ & $z$ & opt & red- & X-ray & X & X & X & X & $f_x\times$ \\
 source & counterpart & & & & & & morph & shift & count & exp & HR1 & HR2 & like & $10^{13}$ \\
 RXS J & SDSS J& & & & & & & & rate & tim & & & & \\
 (1) & (2) & (3) & (4) & (5) & (6) & (7) & (8) & (9) & (10) & (11) & (12) & (13) & (14) & (15)\\
\tableline
002200.9+000659 & 002200.95+000657.9 & 20.48 & 20.03 & 19.28 & 18.87 & 18.55 & 3 & 0.306 & 0.0973 & 403 & 0.08 & 0.12 & 55 & 9.05 \\
003514.9+151513 & 003514.72+151504.1 & 17.32 & 16.93 & 16.59 & 16.31 & 16.05 & 6 & 1.090 & 0.2358 & 359 & 0.35 & 0.25 & 235 & 26.95 \\
005041.0$-$092855 & 005041.31$-$092905.1 & 16.78 & 16.30 & 16.03 & 15.77 & 15.57 & 6 & ... & 0.1974 & 536 & 0.06 & 0.16 & 256 & 21.34 \\
005620.1$-$093626 & 005620.07$-$093629.8 & 18.33 & 17.55 & 16.96 & 16.66 & 16.30 & 3 & 0.103 & 0.3558 & 335 & 0.52 & 0.14 & 298 & 38.70 \\
014126.8$-$092857 & 014125.83$-$092843.6 & 18.11 & 17.59 & 17.21 & 16.97 & 16.68 & 6 & 0.500 & 0.0344 & 440 & -0.01 & 0.24 & 25 & 3.20 \\
\tableline
\end{tabular}
\tablenotetext{a}{[The complete version of this table is in the 
electronic
edition of the journal. The printed version contains only a 
sample.]}\end{table}

\begin{table}
\setlength{\tabcolsep}{0.06in}
\tablenum{6}
\caption{Derived Parameters of RASS/SDSS BL~Lac Candidates\tablenotemark{a}}
\begin{tabular}{llrrrrrrrl}
\tableline
\tableline
 RASS X-ray & SDSS optical & $g_o$ & red- & $f_x^c \times$ & $log(L_x)$ & $log(l_{opt})$ & $log(l_{x})$ & $\alpha_{ox}$ & comment \\
 source &  counterpart & & shift & $10^{13}$ & & 2500\AA\ & 2~kev &  & \\
 RXS J & SDSS J & &  & & & & & \\
 (1) & (2) & (3) & (4) & (5) & (6) & (7) & (8) & (9) & (10)\\
\tableline
002200.9+000659 & 002200.95+000657.9 & 19.94 & 0.306 & 21.02 & 44.86 & 28.87 & 26.33 & 0.98 & BL?,radio \\
003514.9+151513 & 003514.72+151504.1 & 16.67 & 1.090 &  73.75 & 46.84 & 31.41 & 28.30 & 1.19 & zunc,radio \\
005041.0$-$092855 & 005041.31$-$092905.1 & 16.18 & ... & 55.50 & ... & ... & ... & 1.39 & zunc,radio,FBQSJ0050-0929 \\
005620.1$-$093626 & 005620.07$-$093629.8 & 17.37 & 0.103 &  101.2 & 44.46 & 28.89 & 25.92 & 1.14 & BL?,radio, \\
014126.8$-$092857 & 014125.83$-$092843.6 & 17.48 & 0.500 & 7.42 & 44.94 & 30.33 & 26.41 & 1.50 & zunc,radio,[HB89]0138-097 \\
\tableline
\end{tabular}
\tablenotetext{a}{[The complete version of this table is in the 
electronic
edition of the journal. The printed edition contains only a 
sample.]}
\end{table}

\end{document}